\documentclass[a4paper,11pt]{article}
\usepackage{graphicx}  
\usepackage{amsmath}  
\usepackage{amssymb}  
\usepackage{bm}
\usepackage{dcolumn}
\usepackage{color}
\usepackage{colortbl}
\usepackage{colordvi}
\usepackage{mathrsfs}
\usepackage{amsfonts}
\usepackage{varioref}
\usepackage{float}
\usepackage{graphicx}
\usepackage{ragged2e}
\usepackage[square, numbers, sort, compress]{natbib}
\usepackage{bibunits}
\usepackage{aas_macros}
\RequirePackage[colorlinks,citecolor=blue,urlcolor=magenta,linkcolor=blue]{hyperref}
\input epsf
\usepackage{tablefootnote}
\usepackage{multirow}
\usepackage[T1]{fontenc}
\usepackage[utf8]{inputenc}
\usepackage{amsmath, amsfonts, amssymb, mathrsfs}
\usepackage{graphicx}
\usepackage{enumitem}
\usepackage{comment}
\usepackage[noabbrev]{cleveref}
\Crefname{subsection}{section}{sections}
\Crefname{subfigure}{figure}{figures}
\Crefname{equation}{ }{ }
\usepackage[utf8]{inputenc}
\usepackage{amsmath, amsfonts, amssymb}
\usepackage{subcaption}
\usepackage{float}
\usepackage{geometry}

\title{A note on methods for computing the critical curve of Kerr-like black holes}
\author{Siddharth Kumar Sahoo \footnote{521ph1007@nitrkl.ac.in} \footnote{siddharth.math.physics@gmail.com}~$^{1}$ and Indrani Banerjee \footnote{banerjeein@nitrkl.ac.in}~$^{1}$\\
{\small{$^{1}$Department of Physics and Astronomy, National Institute of Technology Rourkela, Odisha-769008, India}}}

\date{ }

\makeindex

\begin{document}
\maketitle
\begin{abstract}
This study systematically compares Bardeen’s, de Vries’s, and Grenzebach et al.’s  celestial coordinate definitions of the critical curve ("shadow") of Kerr-like black holes.  We find that all three definitions agree for black holes in vacuum or surrounded by inhomogeneous plasma observed from large distances. However,  they diverge for  observers located at a finite distance: Bardeen’s definition yields the smallest critical curve, while de Vries’s yields the largest. When homogeneous plasma is considered,  critical curve computed using Bardeen’s definition deviates from the other two even at large distances and  contracts compared to the vacuum case with increasing plasma density. This is in clear  contradiction with the behaviour predicted by de Vries’s, Grenzebach et al.’s definitions, and previous gravitational lensing studies.  We derive de Vries’s definition  assuming a critical curve on the observer’s sky plane and explain its discrepancy with Grenzebach et al.’s definition. We further explore the effect of the change of tetrad on the  critical curve.  Using  Bardeen and Carter tetrads, we plot the critical curve for Schwarzschild and Kerr black holes in the presence of plasma, highlighting that tetrad changes  introduce only a horizontal shift in the critical curve. 
\end{abstract}
%\tableofcontents
\section{Introduction\label{Introduction}}
The images of the supermassive black holes M87* and Sgr A*, released by the Event Horizon Telescope (EHT) collaboration \citep{EventHorizonTelescope:2019dse,EventHorizonTelescope:2019ggy,EventHorizonTelescope:2019ths,EventHorizonTelescope:2022wkp,EventHorizonTelescope:2022wok,EventHorizonTelescope:2025dua,EventHorizonTelescope:2026cny}, have opened a new window for studying the near horizon structure of black holes \citep{EventHorizonTelescope:2025vum,EventHorizonTelescope:2019pgp,EventHorizonTelescope:2021bee,EventHorizonTelescope:2021srq,EventHorizonTelescope:2022urf,EventHorizonTelescope:2022exc,EventHorizonTelescope:2023hrl,EventHorizonTelescope:2023gtd,EventHorizonTelescope:2024hpu}, testing general relativity \citep{EventHorizonTelescope:2020qrl,Berti:2015itd,Will:2014kxa,Ferreira:2019xrr,EventHorizonTelescope:2022xqj} and constraining alternate gravity theories \citep{Psaltis:2008bb,EventHorizonTelescope:2021dqv,Islam:2024sph,Sahoo:2023czj,KumarSahoo:2025igt,KumarSahoo:2025leq,Ditta:2023wye,Afrin:2022ztr,Banerjee:2022jog,Vagnozzi:2022moj}. These images revealed a dark central region, known as the shadow, surrounded by a bright ring, referred to as the primary ring. The formation of these features is primarily attributed to the gravitational capture and lensing of light rays  originating from a distant source or emitted from the accretion disc surrounding the black hole \citep{Gralla:2019xty,Falcke:1999pj,Luminet:1998wr}.

The angular sizes and shapes of these images were found to be  consistent  with  that of the critical curve of a   Kerr black hole\citep{Johnson:2019ljv,Perlick:2015vta}. The critical curve (coined in \citep{Gralla:2019xty}), generally referred to as the shadow, is  traced on the observer's sky plane \citep{Gralla:2019xty,Gralla:2020yvo,Perlick:2021aok}.  It is the boundary on the observer's sky plane that maps to all light rays which, when traced backwards from the observer, approach the black hole's unstable spherical photon orbits  or the photon region \citep{Gralla:2019xty,Gralla:2020yvo,Teo:2003ltt,Perlick:2004tq}. For a Schwarzschild black hole the critical curve was initially studied by  Synge and Zel'dovich \& Novikov \citep{Synge:1966okc,Zeldovich1966}, and then Bardeen  computed the critical curve for a Kerr black hole \citep{Bardeen:1973,Chandrasekharbook}. Bardeen's celestial coordinate definitions\footnote{—unless otherwise stated, by "definition" or "approach" we mean the method of defining celestial coordinates of the critical curve—} for the critical curve became a standard for  subsequent studies of the critical curve of Kerr-like black holes \citep{Takahashi:2005hy,Johannsen:2013vgc}. de Vries, in his study of the critical curve for a Kerr-Newman black hole \citep{2000CQGdeVries}, closely followed Bardeen's method but employed a different definition for computing the critical curve.  It is worth mentioning that for an observer at a large distance, Vàzquez and Esteban    \citep{Vazquez:2003zm}  also proposed a method for computing the  critical curve of a Kerr black hole.  The approach of Vàzquez and Esteban    \citep{Vazquez:2003zm} used the  behaviour of the Boyer-Lindquist coordinates at large distances, unlike Bardeen \citep{Bardeen:1973,Chandrasekharbook} and de Vries \citep{2000CQGdeVries}, which used the tetrad momenta components to define the celestial coordinates of the shadow.   Interestingly, the definitions yield identical critical curve for large observer distances, when applied to Kerr-like black holes \citep{Bardeen:1973,Takahashi:2005hy,2000CQGdeVries}. 

Later, Grenzebach et al. \citep{Grenzebach:2014fha,Grenzebach2015book} introduced a method for plotting the critical curve of Kerr-Newman-NUT black holes, which is also applicable to observers at finite distances. This method was shown to produce results consistent with  Bardeen's approach for distant observers in the Kerr case, as demonstrated by Perlick and Tsupko \citep{Perlick:2017fio,Perlick:2021aok}. Grenzebach et al.'s approach further enabled analytical studies of the apparent shapes of shadows for observers at  a finite distance from rotating black holes. Tsupko and Perlick  \citep{Perlick:2017fio} further extended  the study to include pressureless non-magnetised plasma. Additionally, they also obtained a condition for the plasma frequency, such that the Hamilton-Jacobi formalism remains separable while  studying the critical curve of Kerr black holes surrounded by plasma. 

In the paper by Chang and Zhu  \citep{Chang:2020miq}, the authors proposed an alternative method for computing the critical curve by using astrometric variables which is also applicable for an observer at a finite distance.  However,  they have also reported a  mismatch between critical curves computed using  Grenzebach et al.'s and Bardeen's approaches when observers were at a finite distance  \citep{Chang:2020miq,Chang:2020lmg}. Moreover, they also reported a mismatch between their approach and Grenzebach et al's approach. This may have  been due to the use of different tetrads by Chang and Zhu  \citep{Chang:2020miq} and Grenzebach et al. \citep{Grenzebach:2014fha,Grenzebach2015book}. In \citep{Chang:2020miq},  Chang and Zhu used static tetrads, unlike Grenzebach et al.  who used Carter tetrads \citep{Carter:1968rr,1993GReGrO}  although both approaches used stereographic projection in order to obtain the critical curve in the observer's sky  plane. In the  paper \citep{Chang:2020lmg},  Chang and Zhu  also studied the effect of the observer on the critical curve. However,  comparisons with de Vries's method were not performed.

 Our analysis aims to investigate the mismatch highlighted in previous studies \citep{Perlick:2021aok,Chang:2020miq,Chang:2020lmg} and also to incorporate de Vries's method into the comparison.  In this study, we systematically compare the methods of Bardeen, de Vries, and Grenzebach et al. for computing the critical curve for a stationary observer, both in vacuum and in the presence of plasma. Apart from studying the differences between the methods for computing the critical curve, we also study the effect of changes in tetrads (Bardeen and Carter tetrads)  on the critical curve.   The separability condition found by Tsupko and Perlick \citep{Perlick:2017fio}, applicable to Kerr-like spacetimes \citep{Bezdekova:2022gib}, offers an analytical framework for comparing critical curves  in the presence of plasma and varying observer conditions.

The paper is organised as follows: In \cref{light ray geodesics section}, we provide a brief overview of light ray geodesics in non-magnetised, pressureless plasma. \Cref{Comparison of different methods section} presents a systematic comparison of Bardeen's, de Vries', and Grenzebach's approaches and discusses their differences. In \cref{Analysing the critical curve using Bardeen tetrad}, we derive the equation of the critical curve in Bardeen tetrads for Schwarzschild and Kerr black holes, comparing the results with those from \cref{Comparison of different methods section}. In \cref{Analysing the critical curve using Carter tetrad} we obtain the equation of the critical curve in Carter tetrads, examining the impact of tetrad changes on the differences in the critical curve obtained using the three aforesaid approaches. Finally, we summarise the main findings of our work and discuss the implications of our results in  \cref{Conclusion}. Throughout this paper, we use the metric signature $(-,+,+,+)$ and consider $G=c=\hbar=1$.
\section{Brief overview of light ray geodesics in a Kerr-like spacetime in presence of plasma\label{light ray geodesics section}}
We consider an asymptotically flat, stationary, rotating, and axisymmetric Kerr-like\footnote{By Kerr-like spacetime, we mean that the Hamilton-Jacobi equations for geodesics in the spacetime are separable, leading to the existence of a generalised Carter constant \citep{Carter:1968rr,Azreg-Ainou:2014pra}. Also, the metric reduces to the Kerr metric in  GR, when $M(r)=\mathcal{M}$.} black hole spacetime described by the following metric  in Boyer-Lindquist coordinates $(t,r,\theta,\phi)$ \citep{Boyer:1966qh}
\begin{align}
    \label{metric}
    ds^2&= g_{\mu\nu}dx^\mu dx^\nu\\
    &=-\left(1-\frac{2 r M(r)}{\rho^2}\right)dt^2-\frac{4 r M(r) a \sin^2\theta}{\rho^2}dtd\phi+\frac{\rho^2}{\Delta}dr^2+\rho^2d\theta^2+\frac{\mathcal{A}\sin^2\theta}{\rho^2}d\phi^2
\end{align}
where $a$ is the spin, $M(r)$ is the mass function, and 
\begin{align}
\label{definitions}
    \Delta&=r^2+a^2-2 r  M(r)\\
     \Sigma&=\Delta+2  r M(r)\\
    \rho^2&=\Sigma-a^2\sin^2\theta\\
    \mathcal{A}&=\Sigma^2-\Delta a^2 \sin^2\theta 
\end{align}
the contravariant components of the metric are given by:
\begin{align}
    g^{tt}&=-\frac{\mathcal{A}}{\rho^2\Delta},\ g^{t\phi}=\frac{2r M(r)a}{\rho^2\Delta},\ g^{\phi\phi}=\frac{\Delta-a^2\sin^2\theta}{\rho^2\Delta\sin^2\theta}\\
    g^{rr}&=\frac{\Delta}{\rho^2},\ g^{\theta\theta}=\frac{1}{\rho^2}
\end{align}
The metric \cref{metric} has two Killing vectors $\partial_t$ and $\partial_\phi$, which correspond to conserved quantities, i.e, energy $E=-p_t$ and angular momentum $L=p_\phi$, respectively. 

We further consider the situation in which the black hole is surrounded by a non-magnetised, pressureless plasma with plasma frequency $\omega^2_P$.  The conditions for the propagation of light in a plasma, considering its dispersive effects, have been extensively studied in \citep{Perlick:2017fio,Perlick:2015vta,Bezdekova:2022gib}.  The Hamiltonian for light rays propagating in a pressureless non-magnetised plasma with plasma frequency  $\omega^2_p$  is given by \citep{Perlick:2017fio,Bezdekova:2022gib,breuer1980propagation,breuer1981propagation},
\begin{align}
    \label{hamiltonian}
    \mathcal{H}=\frac{1}{2}\left(g^{\mu\nu}p_\mu p_\nu+ \omega^2_p\right)
\end{align}
and for light ray geodesics in plasma $\mathcal{H}=0$ \citep{breuer1980propagation,breuer1981propagation,Perlick:2017fio}. 
In the presence of plasma the Hamilton-Jacobi equation for $\mathcal{H}$ is separable for Kerr-like spacetimes \emph{iff} the plasma frequency satisfies a separability condition as obtained in \citep{Perlick:2017fio,Bezdekova:2022gib}. For the purpose of our present work, we will focus on plasma environments which   satisfy the separability condition \citep{Perlick:2017fio,Bezdekova:2022gib}, and  thus can be written as 
\begin{align}
\label{separable plasma condition}
\omega^2_P=\frac{f(r)+g(\theta)}{\rho^2}    
\end{align}
Note that substituting \cref{separable plasma condition} in $\mathcal{H}$ and using one of Hamilton's equations, 
$\dot{x}^\mu=\frac{\partial\mathcal{H}}{\partial p^\mu}$, we obtain $\dot{x}^\mu=p^\mu$.

The geodesic equations in the presence of plasma can be written in the form \citep{Perlick:2017fio,KumarSahoo:2025igt},
\begin{align}
    \label{unscaled geodesic equations pr}
    \Delta^2 p^2_r&=-(\mathcal{Q}+f(r))+(\Sigma p_t+a p_\phi)^2\\
  \label{unscaled geodesic equations p theta}
    p^2_\theta&=\mathcal{Q}-g(\theta)-\left(\frac{p_\phi}{\sin{\theta}}+a \sin{\theta}   p_t\right)^2\\
    \label{unscaled geodesic equations p t}
        p_t&=-E\\
    \label{unscaled geodesic equations p phi}
    p_\phi&=L
\end{align}
where $E$ is the energy of the photon as measured by an asymptotic observer and $L$ is the angular momentum of the photon.  The quantity $\mathcal{Q}$ is the generalised Carter's constant\citep{Carter:1968rr,Perlick:2017fio}.   For the sake of simplicity, we will consider light rays of  fixed asymptotic frequency $\omega_0=E$. Scaling \cref{unscaled geodesic equations pr,unscaled geodesic equations p theta,unscaled geodesic equations p phi,unscaled geodesic equations p t} with $E$, we obtain
\begin{align}
    \label{scaled geodesic equations V}
    \frac{\Delta^2 p^2_r}{E^2}&=-(\mathcal{\chi}+f(r))+(-\Sigma +a \eta)^2=V(r)\\
    \label{scaled geodesic equations W}
    \frac{p^2_\theta}{E^2}&=\mathcal{\chi}-g(\theta)-\left(\frac{\eta}{\sin{\theta}}-a \sin{\theta}\right)^2=W(\theta)\\
    \label{scaled geodesic equations pt}
    \frac{p_t}{E}&=-1\\
    \label{scaled geodesic equations p phi}
    \frac{p_\phi}{E}&=\eta
\end{align}
 where $\chi=\mathcal{Q}/E^2$ and $f(r)\equiv f(r)/E^2$ and $g(\theta)\equiv g(\theta)/E^2$ \citep{Perlick:2017fio,Grenzebach:2014fha,Grenzebach2015book,Carter:1968rr}. 

For the computation of the critical curve, we need to determine the spherical photon orbits, or more precisely, the photon region containing unstable spherical photon orbits \citep{Teo:2003ltt,Grenzebach:2014fha,Grenzebach2015book,Perlick:2017fio}.  For the existence of spherical photon orbits, the constants $\chi$ and $\eta$ must satisfy the following conditions 
 \begin{align}
 V(r)|_{r=r_{sp}}=0\text{ and }\frac{d\ V(r)}{d r}\Big|_{r=r_{sp}}=0    
 \end{align}
 where $r_{sp}$ corresponds to the Boyer-Lindquist radius of the spherical photon orbits.
Using these conditions in \cref{scaled geodesic equations V}  gives us the impact parameters  $\eta$ and $\chi$ as
\begin{align}
    \label{a eta}
    a\eta&=\Sigma-\frac{\Delta}{\Delta' } \left(\Sigma'+\sqrt{\Sigma'^2-\Delta'f'(r)}\right)\Bigg|_{r=r_{sp}}\\
    \label{chi}
    \chi&=\frac{\left(a\eta-\Sigma\right)^2}{\Delta}-f(r)\Bigg|_{r=r_{sp}}
\end{align}
 It must be noted that  $\chi$ is related to  $\chi_B=Q_B/E^2$,  the specific Carter constant used in  \citep{Carter:1968rr,Teo:2003ltt,2000CQGdeVries,Bardeen:1973,Chandrasekharbook}. The exact  relation between $\chi$ and $\chi_B$  is \citep{2000CQGdeVries}
\begin{align}
\label{chiB and chi}
    \chi_B=\chi-(\eta-a)^2
\end{align}
Substituting \cref{a eta,chi} in \cref{scaled geodesic equations W}  and finding the range of $r_{sp}$ and $\theta$ in which $W(\theta)\geq0$, will give us the photon region \citep{2000CQGdeVries,Perlick:2017fio}. 
The metric \cref{metric} describes the Kerr spacetime when,
\begin{align}
    \label{kerr metric}
    M(r)&=1\\
    \label{kerr metric Delta}
    \Delta&=r^2+a^2-2 r\\
    \label{kerr metric Sigma}
    \Sigma&=r^2+a^2\\
    \label{kerr metric rho}
    \rho^2&=r^2+a^2\cos^2\theta\\
    \label{kerr metric A}
    \mathcal{A}&=(r^2+a^2)^2-\Delta a^2 \sin^2\theta
\end{align}
Note that, throughout the paper we have scaled the  distances with the gravitational radius $r_g=G\mathcal{M}/c^2$, where $\mathcal{M}$ is the ADM mass of the black hole. Substituting \cref{kerr metric Delta,kerr metric Sigma} in \cref{a eta,chi}, we get the constants $\eta$ and $\chi$ for the Kerr black hole \citep{Perlick:2017fio} surrounded by plasma
\begin{align}
    \label{kerr a eta}
     a\eta&=r^2+a^2-\frac{r^2+a^2-2 r}{2 (r-1) } \left(2 r+\sqrt{4 r^2-2 (r-1)f'(r)}\right)\Bigg|_{r=r_{sp}}\\
      \label{kerr chi}
      \chi&=\frac{\left(a\eta-(r^2+a^2)\right)^2}{r^2+a^2-2 r}-f(r)\Bigg|_{r=r_{sp}}
\end{align}

We now proceed to set up the observer and  the frame for computing the critical curve of a Kerr-like black hole. Consider a stationary observer at a  distance $D$ from a Kerr-like black hole.  Let the angle made by the observer's line of sight with the spin axis of the black hole be  $\theta_i$.   We set up a stationary, orthonormal tetrad \citep{1993GReGrO}  $e^{\mu}_{(i)}$ at the location of the observer of the form:

\begin{align}
\label{stationary tetrads et}
    e_{(t)}&=\left(e^t_{(t)},0,0,e^\phi_{(t)}\right)\bigg|_{(r=D,\theta=\theta_i)}\\
    \label{stationary tetrads er}
    e_{(r)}&=\left(0,e^r_{(r)},0,0\right)\bigg|_{(r=D,\theta=\theta_i)}\\
    \label{stationary tetrads e theta}
    e_{(\theta)}&=\left(0,0,e^\theta_{(\theta)},0\right)\bigg|_{(r=D,\theta=\theta_i)}\\
    \label{stationary tetrads e phi}
    e_{(\phi)}&=\left(e^t_{(\phi)},0,0,e^\phi_{(\phi)}\right)\bigg|_{(r=D,\theta=\theta_i)}\\
    \text{where, }e^r_{(r)}&=\frac{\sqrt{\Delta}}{\rho}\text{ and }e^\theta_{(\theta)}=\frac{1}{\rho}\nonumber
\end{align}
For the above orthonormal tetrads, $g_{\mu\nu}e^\mu_{(i)}e^\nu_{(j)}=\xi_{(i)(j)}$,  where $\xi_{(i)(j)}$ is the Minkowski tensor. The local components of the momenta of the light ray that reach the observer can be obtained using the relations $p_{(i)}=p_\mu e^\mu_{(i)}$ or $p^{(i)}=\xi^{(i)(j)}p_{(j)}$, which are calculated at the observer location  $(r=D,\theta=\theta_i)$.  We express the tetrad components $p^{(a)}$ in terms of the coordinate momenta $p_\mu$ of the light ray at the location of the observer as
\begin{align}
    \label{tetrad momenta components pt}
    p^{(t)}&=-p_{(t)}=-e^t_{(t)}p_t-e^\phi_{(t)}p_{\phi}\bigg|_{(r=D,\theta=\theta_i)}\\
    \label{tetrad momenta components pr}
    p^{(r)}&=p_{(r)}=e^r_{(r)}p_r\bigg|_{(r=D,\theta=\theta_i)}\\
   \label{tetrad momenta components p theta}
    p^{(\theta)}&=p_{(\theta)}=e^\theta_{(\theta)}p_{\theta}\bigg|_{(r=D,\theta=\theta_i)}\\
    \label{tetrad momenta components p phi}
    p^{(\phi)}&=p_{(\phi)}=e^t_{(\phi)}p_t+e^\phi_{(\phi)}p_\phi\bigg|_{(r=D,\theta=\theta_i)}
\end{align}
Additionally, as $\mathcal{H}=0$  \citep{Perlick:2017fio,1980RSPSA.370..389B,breuer1981propagation} thus,
\begin{align}
    \label{tetrad momenta norm}
    (p^{(t)})^2-\big\{(p^{(r)})^2+(p^{(\theta)})^2+(p^{(\phi)})^2\big\}=\omega^2_P|_{(r=D,\theta=\theta_i)}
\end{align}
Using \cref{scaled geodesic equations V,scaled geodesic equations W,scaled geodesic equations pt,scaled geodesic equations p phi} in \cref{tetrad momenta components pt,tetrad momenta components pr,tetrad momenta components p theta,tetrad momenta components p phi} we obtain 
\begin{align}
    \label{scaled tetrad pt}
    \frac{p^{(t)}}{E}&=e^t_{(t)}-e^\phi_{(t)}\eta\\
    \label{scaled tetrad pr}
    \frac{p^{(r)}}{E}&=e^r_{(r)}\frac{p_r}{E}=\frac{e^r_{(r)}\ \sqrt{V(r)}}{\Delta}\\
    \label{scaled tetrad ptheta}
    \frac{p^{(\theta)}}{E}&= e^\theta_{(\theta)}\frac{p_\theta}{E}=e^\theta_{(\theta)} \sqrt{W(\theta)}\\
    \label{scaled tetrad p phi}
    \frac{p^{(\phi)}}{E}&=-e^t_{(\phi)}+e^\phi_{(\phi)}\eta
    \end{align}
    The observer's sky plane is the plane perpendicular to the line joining the observer and the black hole. The black hole is the origin of the  coordinate system on the plane, with the $Y$ and $X$ axes  parallel to $-e^\mu_{(\theta)}$  and $e^\mu_{(\phi)}$, respectively \citep{Grenzebach:2014fha,Grenzebach2015book,Perlick:2004tq,Bardeen:1973,Chandrasekharbook}.  To obtain the critical curve, we have to compute the tangent directions in the tetrad frame of the observer. 

\section{Comparison of different methods  for plotting critical curve of a Kerr-like black hole\label{Comparison of different methods section}}
%%%%%%%%%%%%%%%%%%%%
For the case of a Kerr black hole, the critical curve for an observer at an infinite distance and a given inclination angle $\theta_i$ was first computed by Bardeen in  \citep{Bardeen:1973,Chandrasekharbook}.  The equations describing the shape of the critical curve were  obtained in \citep{Bardeen:1973,Chandrasekharbook} using the following definitions
\begin{align}
    \label{finite bardeen chandrasekhar X}
    X_B&= \frac{-r\ p^{(\phi)}}{p^{(t)}}\Bigg|_{(r=D,\theta=\theta_i)}\\
    \label{finite bardeen chandrasekhar Y}
    Y_B&=\frac{r\ p^{(\theta)}}{p^{(t)}}\Bigg|_{(r=D,\theta=\theta_i)}
\end{align}
Let $(X_b,Y_b)$ represent  the parametric equation of the critical curve in the observer's sky plane computed using Bardeen's definition $(X_B,Y_B) $ for an  observer at infinity. Then the equation of the   critical curve $(X_b,Y_b)$ for a Kerr black hole  \citep{Bardeen:1973,Chandrasekharbook}  is  given by:
\begin{align}
    \label{bardeen chandrasekhar X}
    X_b&=\lim_{D\rightarrow\infty} X_{B}=-\frac{\eta}{\sin\theta_i}\\
    \label{bardeen chandrasekhar Y}
    Y_b&=\lim_{D\rightarrow\infty}Y_{B}=\sqrt{\chi_B+a^2 \cos^2{\theta_i}-\eta^2\cot^2\theta_i}
\end{align}
For the case of a Kerr-Newman black hole,  de Vries \citep{2000CQGdeVries} closely followed \citep{Bardeen:1973,Chandrasekharbook} but computed the critical curve using the following definitions
\begin{align}
\label{finite De Vries X}
X_D&=\frac{-r\ p^{(\phi)}}{p^{(r)}}\Bigg|_{(r=D,\theta=\theta_i)}\\
\label{finite De Vries Y}
Y_D&=\frac{r\ p^{(\theta)}}{p^{(r)}}\Bigg|_{(r=D,\theta=\theta_i)}
\end{align}
Let $(X_d,Y_d)$ represent  the parametric equation of the critical curve  in the observer's sky plane for an observer at infinity, computed using the above definitions. When $(X_D,Y_D)$ is used to compute the parametric equation of the critical curve of a Kerr black hole for an observer at infinity, we get the same result as \cref{bardeen chandrasekhar X,bardeen chandrasekhar Y}, i.e,
\begin{align}
    \label{De Vries X}
    X_d&=\lim_{D\rightarrow\infty}X_D=-\frac{\eta}{\sin\theta_i}\\
    \label{De Vries Y}
    Y_d&=\lim_{D\rightarrow\infty}Y_D=\sqrt{\chi_B+a^2 \cos^2{\theta_i}-\eta^2\cot^2\theta_i}
\end{align}

Comparing   \cref{finite bardeen chandrasekhar X} with \cref{finite De Vries X} and  \cref{finite bardeen chandrasekhar Y} with \cref{De Vries Y}, we note that the denominators of $(X_B,Y_B)$ and $(X_D,Y_D)$ are different. We also note that even if  $(X_B,Y_B)$ and $(X_D,Y_D)$ are  not the same by definition,  the equations of the critical curve of  a Kerr black hole  computed using both definitions  for an observer at infinity are consistent \citep{Bardeen:1973,Chandrasekharbook,2000CQGdeVries}.   There are two reasons behind this  consistency. First, both Bardeen \citep{Bardeen:1973,Chandrasekharbook}  and de Vries \citep{2000CQGdeVries} use the locally non-rotating frame tetrad \citep{Bardeen:1972fi} for the stationary observer\footnote{; refer also to the discussion after equation (57) in \citep{Perlick:2017fio} and section IV in \citep{Perlick:2021aok}.}.  Secondly, in a Kerr-like spacetime, for an  observer  at infinity, 
\begin{align}
    \label{tetrad pt pr asymptoic match}
    \lim_{D\rightarrow\infty}\big\{(p^{(t)})^2-(p^{(r)})^2\big\}=0
\end{align}

When condition \cref{tetrad pt pr asymptoic match} is satisfied\footnote{ which is indeed satisfied for Kerr and Kerr-Newman black hole.} but a  stationary orthonormal tetrad  other than the  Bardeen tetrad \citep{Bardeen:1972fi}  is used for an observer at infinity, we  obtain a horizontally shifted critical curve due to a shift in the origin of the observer's sky plane \citep{Perlick:2017fio, Perlick:2021aok,Kobialko:2022ozq,Kobialko:2023qzo}. 
However, if  the condition  \cref{tetrad pt pr asymptoic match} is not satisfied for an observer at infinity,   i.e, 
\begin{align}
               \label{infinite mismatch}
      \lim_{D\rightarrow\infty}\big\{(p^{(t)})^2-(p^{(r)})^2\big\}\neq0
\end{align}
the critical curve computed using $(X_B,Y_B)$ and $(X_D,Y_D)$  will not be the same. This  condition   arises when a Kerr black hole is surrounded by a non-magnetised, pressureless homogeneous plasma. Furthermore, if we are computing the critical curve for an observer at a  finite distance from a Kerr black hole, then $(X_B, Y_B)$ and $(X_D, Y_D)$ will give different equations because,
\begin{align}
    \label{finite mismatch}
      (p^{(t)})^2-(p^{(r)})^2\neq0
\end{align}

\begin{figure}[ht]
    \centering
    \begin{subfigure}{0.45\textwidth}
    \centering
        \includegraphics[scale=0.4]{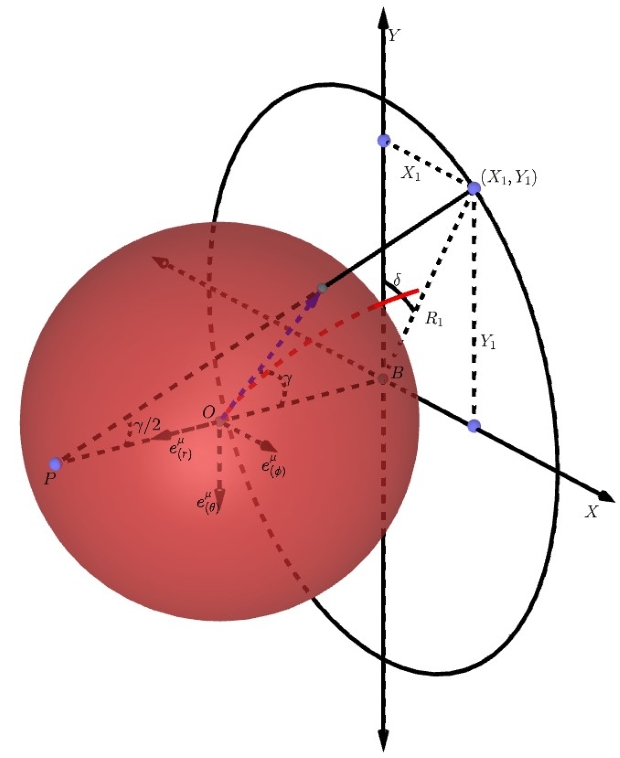}
    \caption{\label{Grenzebach1}}
    \end{subfigure}
        \hfill
        \begin{subfigure}{0.45\textwidth}
        \centering
        \includegraphics[scale=0.4]{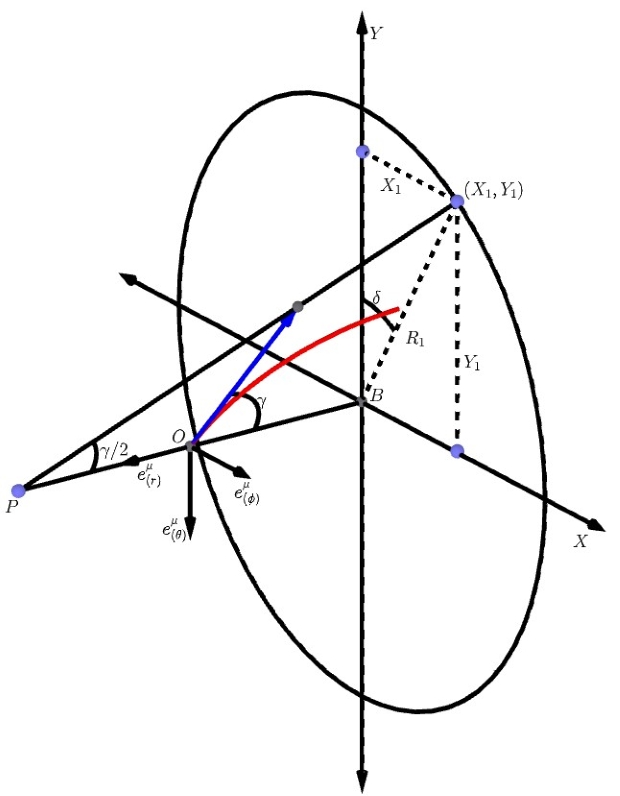}
        \caption{\label{Grenzebach2}}
    \end{subfigure}
    \caption{\label{Grenzebach3D.jpg} A schematic representation of  Grenzebach et al. \citep{Grenzebach:2014fha,Grenzebach2015book} method of plotting the critical curve on the observer's sky plane. $O$ represents the position of the  observer, which is also the centre of the celestial sphere. $P$ is the pole of the celestial sphere, while the other pole  $B$ represents the position of the black hole, which is also the origin of the of the Cartesian coordinate system  on the observer's sky plane. The $X$ and $Y$ axes on the observer's sky plane are parallel and anti-parallel to the $e^{\mu}_{(\phi)}$  and $e^{\mu}_{(\theta)}$, respectively.}
\end{figure}

For an observer at a finite distance, the standard method for plotting the critical curve of a Kerr‑like black hole is the one proposed by Grenzebach et al. \citep{Grenzebach:2014fha,Grenzebach2015book}.  In  Grenzebach et al. \citep{Grenzebach:2014fha,Grenzebach2015book}  approach, the equation of the critical curve of a Kerr-like black hole for  an observer at  finite distance $D$  is  obtained by  using the following definitions :
    \begin{align}
    \label{Grenzebach XG}
    \tilde{X}_G&=2 \tan\left(\frac{\gamma}{2}\right) \sin{\delta}\Bigg|_{(r=D,\theta=\theta_i)}\\
    \label{Grenzebach YG}
    \tilde Y_G&=2 \tan\left(\frac{\gamma}{2}\right) \cos{\delta}\Bigg|_{(r=D,\theta=\theta_i)}
\end{align}
Let the parametric equation of the critical curve  on the observer's sky plane, computed using the above definitions, be $(X_G,Y_G)$\footnote{Throughout our paper, we will plot the critical curve using $(X_G,Y_G)$ while discussing about Grenzebach et al. approach \citep{Grenzebach:2014fha,Grenzebach2015book}. Note that Grenzebach et al.  \citep{Grenzebach:2014fha,Grenzebach2015book}, Tsupko and Perlick's paper \citep{Perlick:2017fio} use dimensionless Cartesian coordinates $(\tilde X_G,\tilde Y_G)$ which can be directly related to $(X_G,Y_G)$ using \cref{Grenzebach X,Grenzebach Y}. }; then 
\begin{align}
    \label{Grenzebach X}
    X_G&= D\ \tilde X_G=2 \ D \tan\left(\frac{\gamma}{2}\right) \sin{\delta}\Bigg|_{(r=D,\theta=\theta_i)}\\
    \label{Grenzebach Y}
    Y_G&= D\ \tilde Y_G=2\ D \tan\left(\frac{\gamma}{2}\right) \cos{\delta}\Bigg|_{(r=D,\theta=\theta_i)}
    \end{align}
In \cref{Grenzebach XG,Grenzebach YG}, $\gamma$ and   $\delta$  are respectively the polar and azimuthal coordinates of the light ray on the celestial sphere, with the observer at the center (refer \cref{Grenzebach3D.jpg}). 
\begin{figure}[ht]
    \centering
        \includegraphics[scale=0.5]{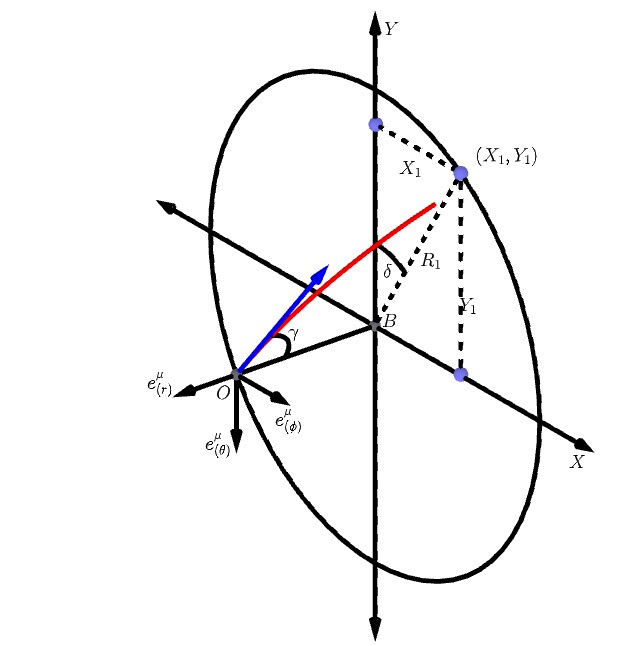}
        \caption{A schematic diagram representing celestial angle   $\gamma$ as the  angle made by the tangent to the light ray with the line of sight of the observer, and  $\delta$   the angle between the  positive $Y$ direction and  the projection of the light ray tangent vector $R_1$ on the  observer's sky plane. }
        \label{gamma _and_delta.png}
    \end{figure}
      
The approach of Grenzebach et al. \citep{Grenzebach:2014fha,Grenzebach2015book}  was later used by  Tsupko and Perlick  \citep{Perlick:2017fio} to compute and analyse the critical curve of a Kerr black hole surrounded by a pressureless, non-magnetised plasma, without assuming an observer at infinity. They  also compared the approach of Grenzebach et al. \citep{Grenzebach:2014fha,Grenzebach2015book} with Bardeen's approach  \citep{Bardeen:1973, Chandrasekharbook}.  The key difference between the two approaches  is that Grenzebach et al. \citep{Grenzebach:2014fha,Grenzebach2015book} uses stereographic projection of the points on the celestial sphere  onto the observer's sky plane, which  represent the tangent vector directions of the light rays originating from the observer;  to obtain the critical curve, which Bardeen's approach \citep{Bardeen:1973, Chandrasekharbook} does not do \citep{Perlick:2021aok}. Other  than this, Grenzebach et al. \citep{Grenzebach:2014fha,Grenzebach2015book} computed the critical curve using the  Carter tetrad \citep{Carter:1968rr} for an observer located at a finite distance from the black hole,  whereas Bardeen \citep{Bardeen:1973, Chandrasekharbook} computed the critical curve using the Bardeen tetrad  \citep{Bardeen:1972fi} for an observer and assumed the observer to be at infinity \citep{Perlick:2021aok}.

We will now obtain general expressions for $(X_B,Y_B)$and  $(X_D,Y_D)$ in terms of the stationary tetrads \cref{stationary tetrads et,stationary tetrads er,stationary tetrads e theta,stationary tetrads e phi}  while also using the null geodesic equations \cref{scaled geodesic equations V,scaled geodesic equations W,scaled geodesic equations p phi,scaled geodesic equations pt}. The definitions  $(X_B,Y_B)$ in \cref{finite bardeen chandrasekhar X,finite bardeen chandrasekhar Y} and $(X_D,Y_D)$ in \cref{finite De Vries X,finite De Vries Y} can be expressed in terms of $V(r),W(\theta)\text{ and }\eta$ as 
    \begin{align}
    \label{XB any tetrad}
    X_B&=\frac{-r\  (-e^t_{(\phi)}+e^\phi_{(\phi)}\eta)}{e^t_{(t)}-e^\phi_{(t)} \eta}\Bigg|_{(r=D,\theta=\theta_i)}\\
    \label{YB any tetrad}
    Y_B&=\frac{r\ e^\theta_{(\theta)}\ \sqrt{W(\theta)})}{e^t_{(t)}-e^\phi_{(t)} \eta}\Bigg|_{(r=D,\theta=\theta_i)}\\
    \nonumber\\
    \label{XD any tetrad}
    X_D&=\frac{-r\ \Delta\  (-e^t_{(\phi)}+e^\phi_{(\phi)}\eta)}{e^r_{(r)}\ \sqrt{V(r)}}\Bigg|_{(r=D,\theta=\theta_i)}\\
    \label{YD any tetrad}
    Y_D&=\frac{r \ \Delta\ \ e^\theta_{(\theta)}\ \sqrt{W(\theta)})}{e^r_{(r)}\ \sqrt{V(r)}}\Bigg|_{(r=D,\theta=\theta_i)}
    \end{align}
    \subsection{Comparison of the three approaches in terms of celestial angles}
       It is also possible to express   $\gamma\text{ and }\delta$ in \cref{Grenzebach XG,Grenzebach YG} in terms of $p^{(a)}$, and vice versa. For this, we first expand the tangent vector $\Gamma^\mu$ of the light ray reaching the observer in terms of the tetrad components $p^{(a)}$  as
    \begin{align}
    \label{tetrad expansion DeVreis}    
    \Gamma^\mu&=-p^{(t)} e^\mu_{(t)}+p^{(\phi)} e^\mu_{(\phi)}+p^{(\theta)} e^\mu_{(\theta)}+p^{(r)} e^\mu_{(r)}\bigg|_{(r=D,\theta=\theta_i)}
   \end{align}
    The celestial angles  $\gamma$  and $\delta$ representing the direction of  $\Gamma^\mu$  may be equivalently interpreted as follows, $\gamma$  is the angle made by the tangent to the light ray with the line of sight of the observer  and $\delta$  represents the angle between the  positive $Y$ direction and  the projection of the light ray tangent vector on the  observer's sky (XY-plane) plane {(refer \cref{gamma _and_delta.png})}.   As done in   \citep{Grenzebach:2014fha,Grenzebach2015book,Perlick:2017fio},  $\Gamma^\mu$ may also be expanded  as  
    \begin{align}
    \label{tetrad expansion grenzebach}    
    \Gamma^\mu&=-\alpha e^\mu_{(t)}+\beta\left(\sin\gamma \sin\delta e^\mu_{(\phi)}+\sin\gamma \cos\delta e^\mu_{(\theta)}+\cos\gamma  e^\mu_{(r)}\right)\bigg|_{(r=D,\theta=\theta_i)}
    \end{align}
    Comparing \cref{tetrad expansion DeVreis,tetrad expansion grenzebach} and using some trigonometry, we obtain
    \begin{align}
     \label{alpha}
    \alpha&=p^{(t)}|_{(r=D,\theta=\theta_i)}\\
    \label{beta}
    \beta&=\sqrt{(p^{(r)})^2+(p^{(\theta)})^2+(p^{(\phi)})^2}\bigg|_{(r=D,\theta=\theta_i)}\\
    \nonumber\\
    \label{gamma}
    \gamma&=\arctan\left(\frac{\sqrt{(p^{(\theta)})^2+(p^{(\phi)})^2}}{p^{(r)}}\right)\Bigg|_{(r=D,\theta=\theta_i)}\\
    \label{delta}
    \delta&=\arctan\left(\frac{-p^{(\phi)}}{p^{(\theta)}}\right)\bigg|_{(r=D,\theta=\theta_i)}
    \end{align}
Additionally, as $\mathcal{H}=0$ for light rays in plasma considered in the present work,    we also obtain
\begin{align}
\label{alpha beta omega}
    \alpha^2-\beta^2|_{(r=D,\theta=\theta_i)}=\omega^2_P|_{(r=D,\theta=\theta_i)}
\end{align}
    
Using the above results, $(X_D,Y_D)$ and $(X_B,Y_B)$ can also be expressed in terms of $\gamma$ and $\delta$.  This will allow us to directly compare  $(X_D,Y_D)$ and $(X_B,Y_B)$ with $(X_G,Y_G)$.   In order to  express $(X_D, Y_D)$ in terms of $\gamma\text{ and }\delta$,  we  invert  \cref{gamma,delta} and get
\begin{align}
    \label{tan x}
    \frac{p^{(\phi)}}{p^{(r)}}&=-\tan\gamma\sin\delta\\
    \label{tan y}
    \frac{p^{(\theta)}}{p^{(r)}}&=\tan\gamma\cos\delta,
\end{align}
these relations are also evident from \cref{gamma _and_delta.png}.
Substituting the above results in \cref{finite De Vries X,finite De Vries Y} we obtain
\begin{align}
    \label{finite XD in gamma delta}
    X_D=D \tan\gamma\sin\delta\\
    \label{finite YD in gamma delta}
    Y_D=D \tan\gamma\cos\delta
\end{align}
Proceeding similarly for Bardeen's definition $(X_B,Y_B)$ we obtain
\footnote{
Rearranging  $\alpha^2-\beta^2=\omega^2_P$,  we can obtain  $ \frac{\beta}{\alpha}=\sqrt{1-\frac{\omega^2_P}{\alpha^2}}$.
}
\begin{align}
    \label{finite XB gamma delta}
    X_B&=\left(\sqrt{1-\frac{\omega^2_P}{\alpha^2}}\right)D\sin\gamma \sin\delta\\
    \label{finite YB gamma delta}
    Y_B&=\left(\sqrt{1-\frac{\omega^2_P}{\alpha^2}}\right)D\sin\gamma \cos\delta
\end{align}

Comparing $(X_G,Y_G)$ in  \cref{Grenzebach X,Grenzebach Y}  with $(X_D,Y_D)$  in   \cref{finite XD in gamma delta,finite YD in gamma delta}, we conclude that for an observer at a finite distance,  the definitions by Grenzebach et al. \citep{Grenzebach:2014fha,Grenzebach2015book}  and de Vries\citep{2000CQGdeVries}  also do not match, and thus will not produce the same critical curve in the observer's sky plane. Thus, in general, for an observer at a finite distance from a Kerr-like black hole, the definitions $(X_B,Y_B)$, $(X_D,Y_D)$, and $(X_G,Y_G)$ do not provide a consistent description of the critical curve on the observer's sky plane. In fact, as $\tan\gamma>2 \tan(\gamma/2)>\sin\gamma$,  $(X_B,Y_B)$ will always be smaller in size compared to $(X_G,Y_G)$, which will be smaller compared to  $(X_D,Y_D)$.  

When  the observer  is  at a large distance,  $\sin\gamma\approx\tan\gamma\approx2\tan(\gamma/2)\approx\gamma$\footnote{refer \cref{Grenzebach3D.jpg} for visualisation.}. Thus,   \cref{finite XD in gamma delta,finite YD in gamma delta,finite XB gamma delta,finite YB gamma delta,Grenzebach X,Grenzebach Y} becomes,
\begin{align}
    \label{large D XB}
    X_B&=\left(\sqrt{1-\frac{\omega^2_P}{\alpha^2}}\right) D\ \gamma
\sin\delta\\
\label{large D YB}
    Y_B&=\left(\sqrt{1-\frac{\omega^2_P}{\alpha^2}}\right) D\ \gamma
\cos\delta\\
\label{large D XD}
    X_D&= D\ \gamma\sin\delta\\
\label{large D YD}
    Y_D&= D\ \gamma\cos\delta\\
    \label{large D XG}
    X_{G}&=D\  \tilde X_G= D\ \gamma\sin\delta\\
\label{large D YG}
    Y_{G}&= D\ \tilde Y_G=D\ \gamma\cos\delta
\end{align}
Clearly,  when  the  observer is at a  large distance and  there is no plasma  $(\omega^2_P=0)$, the above equations produce the same critical curve. Interestingly, for the case of homogeneous plasma $(\omega^2_P=const.)$, $(X_B,Y_B)$  predicts a decrease in shadow size as $\omega^2_P$ increases. This is in clear contradiction with results reported in the literature, which show that homogeneous plasma has an increasing effect on the shadow size or, more precisely, the size of the critical curve  \citep{Chowdhuri:2020ipb,Perlick:2017fio,KumarSahoo:2025igt,KumarSahoo:2025leq,Molla:2022izk}. Additionally, gravitational lensing studies have also reported on the magnifying effect of homogeneous plasma on lensing images  \citep{Er:2013efa,Tsupko:2013cqa,Bisnovatyi-Kogan:2010flt}.  This means that the  definitions $(X_B,Y_B)$ used by Bardeen \citep{Bardeen:1973,Chandrasekharbook} do not seem to be correct.
\newpage

\subsection{Derivation of de Vries definition from the  critical curve\label{Derivation of de Vries definition from the  critical curve}}  
 \begin{figure}[ht]
    \begin{subfigure}{0.5\textwidth}
        \includegraphics[width=\textwidth,height=\textwidth]{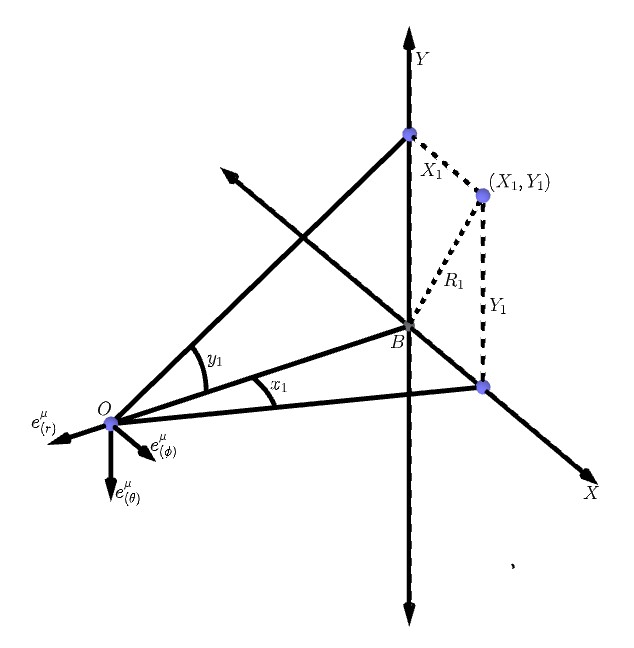}
        \caption{\label{angular_height _and_width.png}}
    \end{subfigure}
    \hfill
        \begin{subfigure}{0.5\textwidth}
          \includegraphics[width=\textwidth,height=\textwidth]{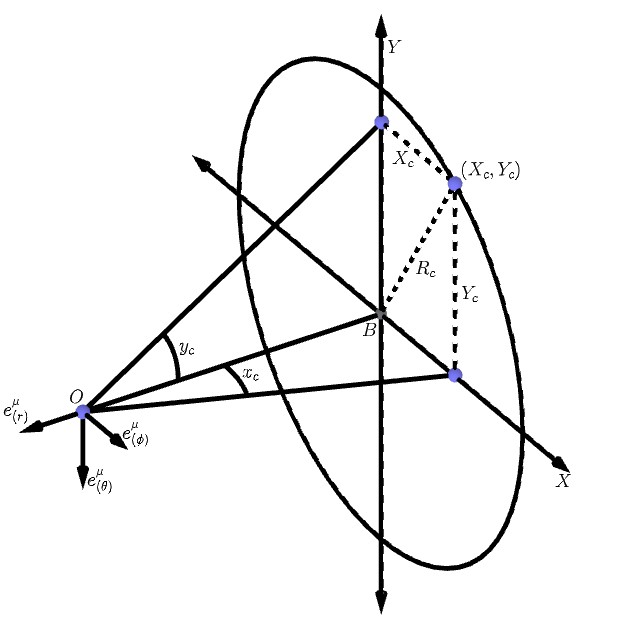}
        \caption{\label{DeVries.png}}
    \end{subfigure}
    
    \caption{In \cref{DeVries.png,angular_height _and_width.png},  $O$ represents the position of the  observer and  $B$ represents the position of the black hole. $B$ is the origin of the of the Cartesian coordinate system  on the observer's sky plane. The $X$ and $Y$ axes are parallel and anti-parallel to the $e^{\mu}_{(\phi)}$  and $e^{\mu}_{(\theta)}$, respectively. In   \Cref{angular_height _and_width.png},   the angular coordinates $(x_1,y_1)$ corresponding to the point $(X_1,Y_1)$ are shown. In  \Cref{DeVries.png}   the angular coordinates $(x_c,y_c)$ corresponding to the point $(X_c,Y_c)$ on the critical curve are shown. \label{Fig3}}
\end{figure}

We have clearly shown the contradictory results that the Bardeen's definition  produces when homogeneous plasma is considered, which can persist even at  large distances. This leaves us with two definitions, $(X_D,Y_D)$ by  de Vries \citep{2000CQGdeVries} and $(X_G,Y_G)$ by Grenzebach et al. \citep{Grenzebach:2014fha} which give the same result at large distances but are inconsistent at distances comparable to the horizon radius or critical curve size  of the black hole. In order to resolve this mismatch, we assume a critical curve of a Kerr-like black hole in the observer's sky plane for a stationary observer {at an inclination $\theta_i$} and at a distance $D$ from the black hole.  We will obtain the expressions for computing {any point $(X_c,Y_c)$ on} the critical curve in terms of $p^{(a)}$.  {This means that the line joining  $(X_c,Y_c)$ (\cref{Fig3}) to the observer location  $O$  represents the tangent to a light ray  originating from the observer in  it's local frame. Such a ray when traced backwards, approaches the unstable spherical photon orbits.   }The angular measures\footnote{$x_1$ and $y_1$   correspond to the vertical angular measure and  the horizontal angular measure of the point $(X_1,Y_1)$ at the location of the observer, respectively.}  $x_1\text{ and }y_1$  of any point $(X_1,Y_1)$ on the observer's sky plane, as measured from the line joining the black hole and the observer, can be obtained using the following relations, 
\begin{align}
    \label{angular width x}
    x_1&=\arctan{\left(\frac{X_1}{D}\right)}\\
    \label{angular height y}
    y_1&=\arctan{\left(\frac{Y_1}{D}\right)}
\end{align}
A visual representation of angular coordinates is shown in   \cref{angular_height _and_width.png}.   Only when the observer is at a large distance, i.e,  $D>>X_1\text{ and }D>>Y_1$,    $x_1$ and $y_1$ may be approximated as
\begin{align}
    \label{angular width large distance}
    x_1&=\frac{X_1}{D}\\
    \label{angular  height large distance}
    y_1&=\frac{Y_1}{D}
\end{align}

Let $(X_c,Y_c)$ be a point on the critical curve  in the observer's sky plane with angular width  $x_c$ and angular height  $y_c$, for an observer located at  $(D,\theta_i)$. Then, $X_c\text{ and }Y_c$ can be expressed in terms of the corresponding angular measures  $x_c\text{ and }y_c$ as (refer \cref{DeVries.png}),  
\begin{align}
    \label{Xc in terms of  xc}
    X_c&=D\tan{x_c}\\
    \label{Yc in terms of  yc}
    Y_c&=D\tan{y_c}
\end{align}
Note  $p^{(\phi)}$,  $ p^{(\theta)}\text{ and }p^{(r)}$ are  the projections of $\Gamma^\mu$ along $e^\mu_{(\phi)}$, $e^\mu_{(\theta)}$  and $e^\mu_{(r)}$. Thus, for any point $(X_c,Y_c)$ with angular measures $x_c\text{ and }y_c$,
\begin{align}
    \label{tan xc}
    \tan{x_c}&=\frac{p^{(\phi)}}{-p^{(r)}}=\frac{X_c}{D}\\
    \label{tan yc}
    \tan{y_c}&=\frac{p^{(\theta)}}{p^{(r)}}=\frac{Y_c}{D}  
\end{align}

Rearranging  \cref{tan xc,tan yc}  we get
\begin{align}
    X_c&=\frac{-D\  p^{(\phi)}}{p^{(r)}}=\frac{-r\  p^{(\phi)}}{p^{(r)}}\Bigg|_{(r=D,\theta=\theta_i)}=X_D\\
    Y_c&=\frac{D\  p^{(\theta)}}{p^{(r)}}=\frac{r\  p^{(\theta)}}{p^{(r)}}\Bigg|_{(r=D,\theta=\theta_i)}=Y_D
\end{align}

Thus, we have shown that the definition  $(X_D,Y_D)$  by de Vries \citep{2000CQGdeVries}  { can be derived by assuming a}    critical curve in the observer's sky plane, in a stationary frame, for a stationary observer at any distance outside the outer horizon of a Kerr-like black hole.  It must be noted that,  although $(X_G,Y_G)$ does not correctly plot the critical curve in the observer's sky plane, the celestial angles $\gamma$ and $\delta $ computed in \citep{Grenzebach:2014fha,Grenzebach2015book,Perlick:2017fio}  correctly trace the {tangent directions corresponding to the critical curve} on the celestial sphere.  The main issue with Grenzelbach et al. approach \citep{Grenzebach:2014fha,Grenzebach2015book} is their use of stereographic projection to obtain the critical curve on the observer sky plane.

{The  reason the  stereographic projection method does  not correctly trace the critical curve on the observer's sky plane at finite distance is the following; the point obtained from the stereographic projection method is not the same as that of the point obtained by extending the light ray tangent vector from the observer location to the observer's sky plane (refer \cref{Grenzebach3D.jpg}). When the observer is at a large distance, the line joining the pole and the stereographic projection point on the observer's sky plane begins to align with the tangent ray direction and thus, the $(X_G,Y_G)$ approximately coincides with $(X_D,Y_D)$.

We conclude this section by giving the  expression for the angular diameter  of the critical curve for an observer at a finite distance. For Kerr-like black holes, the critical curve is symmetric about the $X-$axis in the observer's sky plane \citep{Bardeen:1973,Chandrasekharbook,2000CQGdeVries,Perlick:2017fio,Perlick:2021aok,Grenzebach:2014fha,Grenzebach2015book}.  
Thus, the vertical angular diameter of the critical curve $\Delta\zeta$ for an observer at distance $D$ is given by,
\begin{align}
\label{delta zeta}
    \Delta\zeta&=2\ y_{c\ max}=2\arctan{\left(\frac{Y_{c\ max}}{D}\right)}
\end{align}
In the above expressions, $Y_{c\ max}$ is the maximum height of the critical curve, and $y_{c\ max}$ is the maximum angular height of the critical curve. 
 
\section{Analysing the critical curve of black holes using Bardeen tetrad\label{Analysing the critical curve using Bardeen tetrad}} 
In the present section, we will study the inconsistencies between $(X_B,Y_B)$, $(X_D,Y_D)$ and $(X_G,Y_G)$ using Bardeen tetrads \citep{Bardeen:1972fi,Bardeen:1973,1993GReGrO}.   We will obtain the parametric equations  of the critical curve in the observer's sky plane for the case of a Kerr-like black hole surrounded by  a non-magnetised, pressureless plasma, using definitions $(X_B,Y_B)$ in \cref{XB any tetrad,YB any tetrad}, $(X_D,Y_D)$ \cref{XD any tetrad,YD any tetrad} and $(X_G,Y_G)$  in \cref{Grenzebach XG,Grenzebach YG}. However, we will focus our analysis on the case of a Schwarzschild and a Kerr black hole.

We will  consider the following plasma profiles \citep{Perlick:2017fio,shapiro1974accretion}
\begin{align}
\label{homogeneous plasma general}
    \text{Profile 1: }\frac{\omega^2_P}{\omega^2_0}&=\alpha_1,\ \text{where,}\ f(r)=\alpha_1 \Sigma \text{ and }\ g(\theta)=-\alpha_1 a^2 \sin^2\theta\\
    \label{inhomogeneous plasma general}
    \text{Profile 2: }\frac{\omega^2_P}{\omega^2_0}&=\frac{\alpha_2 \sqrt{r}}{\rho^2},\ \text{where,}\ f(r)=\alpha_2 \sqrt{r} \text{ and }\ g(\theta)=0
\end{align}
 where $\omega_P^2$ in given in \cref{separable plasma condition}. 
Profile 1 described in \cref{homogeneous plasma general} represents homogeneous plasma for which $\omega^2_P=const.$ Note that for homogeneous 
plasma, the necessary and sufficient condition is  $f(r)+g(\theta)=\rho^2$, and therefore the choice of $f(r)$ and $g(\theta)$  is not unique\footnote{Thus, $f(r)=\alpha_1 r^2$ and $g(\theta)=\alpha_1 a^2\cos^2\theta$ is equivalent to the choice $f(r)=r^2+a^2$ and $g(\theta)=-\alpha_1 a^2 \sin^2\theta$.}. Profile 2 described in \cref{inhomogeneous plasma general} behaves like the Shapiro plasma profile \citep{shapiro1974accretion} at a large distance and is an example of an inhomogeneous plasma profile. Furthermore, $\alpha_1$ and $\alpha_2$ are dimensionless constants whose maximum value is determined by the light propagation condition in plasma \citep{KumarSahoo:2025igt,KumarSahoo:2025leq,Perlick:2017fio}.
For the case of a Kerr black hole using \cref{kerr metric Sigma,kerr metric rho} we get,
\begin{align}
    \label{kerr homogeneous}
    \text{Profile 1: }\frac{\omega^2_P}{\omega^2_0}&=\frac{\alpha_1(r^2+a^2)-\alpha_1 a^2 \sin^2\theta}{r^2+a^2\cos^2\theta}\text{, where $0\leq\alpha_1\leq1$.}\\
    \label{kerr inhomogeneous}
    \text{Profile 2: }\frac{\omega^2_P}{\omega^2_0}&=\frac{\alpha_2 \sqrt{r}}{r^2+a^2\cos^2\theta}\text{, where $0\leq\alpha_2\lesssim14$.}
\end{align}

For an observer at a finite distance $D$ and inclination angle $\theta_i$, the Bardeen tetrads \citep{1993GReGrO,Bardeen:1972fi,Bardeen:1973} are  given by,
\begin{align}
 \label{Bardeen tetrad et}
    e_{(t)}&=\left( {\frac{ 1}{\rho}\sqrt\frac{\mathcal{A}}{{\Delta}}},0,0,\frac{2 r M(r)a}{\sqrt{\mathcal{A}\rho^2\Delta}}\right)\Bigg|_{(r=D,\theta=\theta_i)}\\
    \label{Bardeen er}
    e_{(r)}&=\left(0,\frac{\sqrt{\Delta}}{\rho},0,0\right)\Bigg|_{(r=D,\theta=\theta_i)}\\
    \label{Bardeen e theta}
    e_{(\theta)}&=\left(0,0,\frac{1}{\rho},0\right)\Bigg|_{(r=D,\theta=\theta_i)}\\
    \label{Bardeen e phi}
    e_{(\phi)}&=\left( 0,0,0,\frac{\rho}{\sqrt{\mathcal{A}}\ \sin\theta}\right)\Bigg|_{(r=D,\theta=\theta_i)}
\end{align}
Using  \cref{Bardeen tetrad et,Bardeen er,Bardeen e theta,Bardeen e phi}  in \cref{scaled tetrad pt,scaled tetrad pr,scaled tetrad ptheta,scaled tetrad p phi,gamma,delta} we get,
\begin{align}
    \label{Bardeen pt}
    \frac{p^{(t)}}{E}&=\frac{1}{\sqrt{\mathcal{A}\rho^2\Delta}}\left(\mathcal{A}-2 r M(r) a\eta\right)\Bigg|_{(r=D,\theta=\theta_i)}\\
    \label{Bardeen pr}
    \frac{p^{(r)}}{E}&=\sqrt{\frac{(\Sigma-a\eta)^2-(\chi+f(r))\Delta}{\rho^2\Delta}}\Bigg|_{(r=D,\theta=\theta_i)}\\
    \label{Bardeen p theta}
    \frac{p^{(\theta)}}{E}&=\sqrt{\frac{1}{\rho^2}\left(\chi-g(\theta)-\left(\frac{\eta}{\sin\theta}-a\sin\theta\right)^2\right)}\Bigg|_{(r=D,\theta=\theta_i)}\\
    \label{Bardeen p phi}
    \frac{p^{(\phi)}}{E}&=\frac{\rho}{\sqrt{\mathcal{A}}}\left(\frac{\eta}{\sin\theta}\right)\Bigg|_{(r=D,\theta=\theta_i)}\\
    \label{Bardeen gamma}
    \gamma&=\arctan{\left(\sqrt{\frac{\Delta}{V(r)}\left(\chi-g(\theta)-\left(\frac{\eta}{\sin\theta}-a\sin\theta\right)^2+\frac{\rho^4\eta^2}{\mathcal{A}\sin^2{\theta}}\right)}\right)}\Bigg|_{(r=D,\theta=\theta_i)}\\
    \label{Bardeen delta}
    \delta&=\arctan{\left({\frac{-\sin{\theta}}{\rho^2\eta}}\sqrt{\mathcal{A}\left(\chi-g(\theta)-\left(\frac{\eta}{\sin\theta}-a\sin\theta\right)^2\right)}\right)}\Bigg|_{(r=D,\theta=\theta_i)}
\end{align}
Using   \cref{Bardeen pt,Bardeen pr,Bardeen p theta,Bardeen p phi}  in  Bardeen's definition $(X_B,Y_B)$ in \cref{finite bardeen chandrasekhar X,finite bardeen chandrasekhar Y}  we get,
\begin{align}
    \label{Bardeen tetrad XB}
  X_B&=\frac{-r\sqrt{\Delta}\ \rho^2\eta\csc\theta}{\mathcal{A}-2 r M(r)a\eta}\Bigg|_{(r=D,\theta=\theta_i)}\\
  \label{Bardeen tetrad YB}
  Y_B&=\frac{r\sqrt{\mathcal{A}\Delta\left(\chi-g(\theta)-(\eta\csc\theta-a\sin\theta)^2\right)}}{\mathcal{A}-2 r M(r) a\eta}\Bigg|_{(r=D,\theta=\theta_i)}
\end{align}

Proceeding similarly for the de Vries definition  $(X_D,Y_D)$ in \cref{finite De Vries X,finite De Vries Y},
\begin{align}
    \label{Bardeen tetrad XD}
    X_D&=\frac{-r\rho^2\sqrt{\Delta}\ \eta\csc\theta}{\sqrt{\mathcal{A}V(r)}}\Bigg|_{(r=D,\theta=\theta_i)}\\
    \label{Bardeen tetrad YD}
    Y_D&=r\sqrt{\frac{\Delta W(\theta)}{V(r)}}\Bigg|_{(r=D,\theta=\theta_i)}
\end{align}
And finally, the Grenzebach et al  definition $(X_G, Y_G)$ becomes,
\begin{align}
    \label{Bardeen tetrad XG}
    X_G&=D\ \tilde X_G=2 D \tan\left(\frac{\gamma}{2}\right) \sin\delta\\
    \label{Bardeen tetrad YG}
    Y_G&=D\ \tilde Y_G= 2 D \tan\left(\frac{\gamma}{2}\right) \cos\delta
\end{align}
where $\gamma$ and $\delta$ are given by \cref{Bardeen gamma,Bardeen delta}. Note that in the above equations, $\chi$ and $\eta $ are functions of $r_{sp}$ as given in   \cref{a eta,chi}, where $r_{sp}$ varies in the range satisfied by the inequality $W(\theta_i)\geq0$.
This makes $(X_B,Y_B)$, $(X_D,Y_D)$ and $(X_G,Y_G)$ parametric functions, with $r_{sp}$ as a parameter. We may plot the critical curve on the observer's sky plane by varying $r_{sp}$ in the allowed range obtained from the inequality
\begin{align}
    \label{rsp range}
    \chi(r_{sp})-g(\theta_i)-\left(\frac{\eta(r_{sp})}{\sin\theta_i}-a\sin\theta_i\right)^2\geq0
\end{align}

For an observer located at infinity (or $D$ is large), the parametric equations of the  critical curves  computed using $(X_B,Y_B)$, $(X_D,Y_D)$ and $(X_G,Y_G)$ are given by
 \begin{align}
     \label{Xb}
     X_b&=\lim_{D\rightarrow\infty}X_B=\frac{-\eta}{\sin\theta_i}\\
     \label{Yb}
     Y_b&=\lim_{D\rightarrow\infty}Y_B=\sqrt{\chi-g(\theta_i)-\left(\eta\csc\theta_i-a\sin\theta_i\right)^2}\\
     \label{Xd}
     X_d&=\lim_{D\rightarrow\infty}X_D=\frac{-\eta\csc\theta_i}{\sqrt{1-\lambda}}\\
     \label{Yd}
     Y_d&=\lim_{D\rightarrow\infty}Y_D=\frac{\sqrt{\chi-g(\theta_i)-\left(\eta\csc\theta_i-a\sin\theta_i\right)^2}}{\sqrt{1-\lambda}}\\
     &\text{where, }\lambda=\lim_{D\rightarrow\infty}\frac{f(r)}{r^2}\Bigg|_{r=D}\nonumber\\
     \label{Xg}
     X_g&=\lim_{D\rightarrow\infty}X_G=X_d\\
     \label{Yg}
     Y_g&=\lim_{D\rightarrow\infty}Y_G=Y_d      
     \end{align}
In the above equations\footnote{As we are considering Kerr-like black holes, when $r$ is large, $\Sigma\sim\Delta \sim r^2$.},   \cref{Xg,Yg} directly follows from the analysis for the observer at a large distance in the  previous section; refer \cref{large D XD,large D YD,large D XG,large D YG}.  Notice that for inhomogeneous plasma with a density decreasing with distance from the black hole, $\lambda\rightarrow0$ and thus, $(X_b,Y_b)$=$(X_d,Y_d)$=$(X_g,Y_g)$. 

However, in the case of homogeneous plasma, using \cref{kerr homogeneous}, we obtain $\lambda\neq0$. Thus, comparing \cref{Xb,Yb,Xd,Yd,Xg,Yg} we get $(X_d,Y_d)=(X_g,Y_g)\neq(X_b,Y_b)$.  Thus, the consistency of Grenzebach et al. \citep{Grenzebach:2014fha,Grenzebach2015book} approach with that of Bardeen's approach \citep{Bardeen:1973} in the large distance limit, as commented by Tsupko and Perlick in \citep{Perlick:2017fio}, does not hold when homogeneous plasma is considered. However, this consistency holds when Grenzebach et al. \citep{Grenzebach:2014fha,Grenzebach2015book} is compared with de Vries's approach \citep{2025CQGHioki}. This has also been shown explicitly in the previous section.

In \cref{Schwarzschild plasma,Kerr plasma}, the critical curve  calculated 
using $(X_B,Y_B)$ in \cref{Bardeen tetrad XB,Bardeen tetrad YB},  $(X_D,Y_D)$ in \cref{Bardeen tetrad XD,Bardeen tetrad YD},  and  $(X_G,Y_G)$ in \cref{Bardeen tetrad XG,Bardeen tetrad YG} for a Schwarzschild and a near extremal Kerr black hole are shown, respectively. Furthermore, in both figures,  the first, second, and third rows correspond to the critical curve plotted in  absence of plasma ($\alpha_1=\alpha_2=0$), in the presence of inhomogeneous plasma  and in the presence of a homogeneous plasma. For all the figures, the observer is in the equatorial plane. As we go from left to right in a row, we can see the effect of decreasing distance on  $(X_B,Y_B)$, $(X_D,Y_D)$ and $(X_G,Y_G)$. As we go from top to bottom in a column, we can see the effect of  plasma parameter  on  $(X_B,Y_B)$, $(X_D,Y_D)$ and $(X_G,Y_G)$ for inhomogeneous and homogeneous plasma. We report the following observations:
\begin{figure}[t!]
    \begin{subfigure}{0.25\textwidth}
    \includegraphics[width=\linewidth,height=\textwidth]{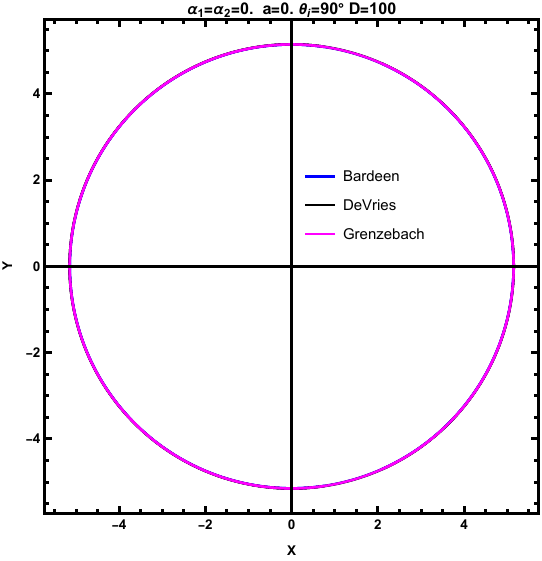}
        \caption{\label{geometric_alpha_0._a_0._plasma_homogeneous_D_100_theta_90.pdf}}
    \end{subfigure}
    \hfill
    \begin{subfigure}{0.25\textwidth}
    \includegraphics[width=\linewidth,height=\textwidth]{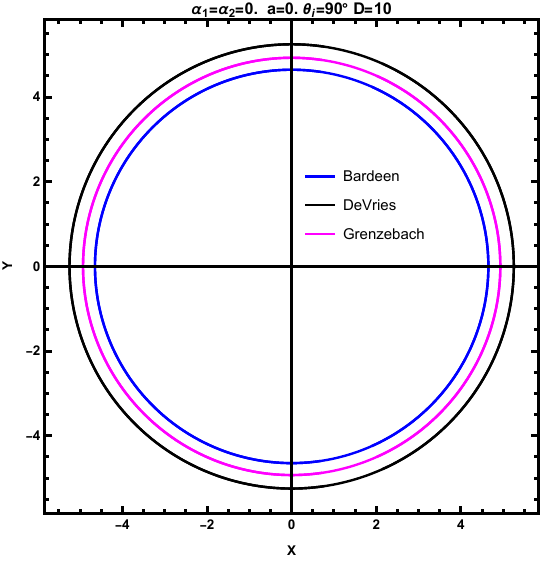}
        \caption{\label{geometric_alpha_0._a_0._plasma_homogeneous_D_10_theta_90.pdf}}
    \end{subfigure}
    \hfill
    \begin{subfigure}{0.25\textwidth}
    \includegraphics[width=\linewidth,height=\textwidth]{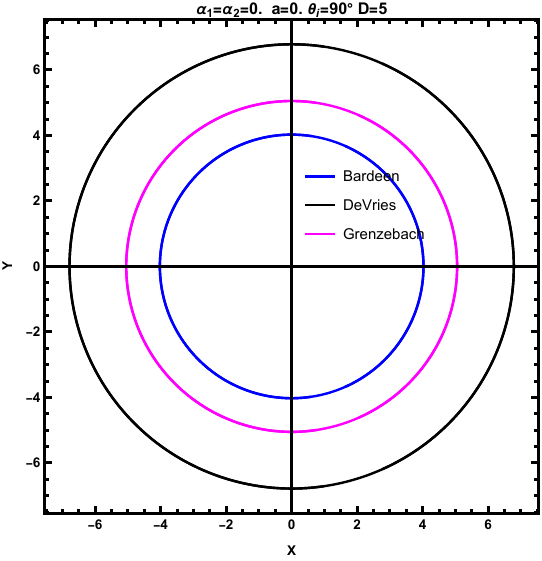}
        \caption{\label{geometric_alpha_0._a_0._plasma_homogeneous_D_5_theta_90.pdf}}
    \end{subfigure}
    \hfill
    \begin{subfigure}{0.25\textwidth}
    \includegraphics[width=\linewidth,height=\textwidth]{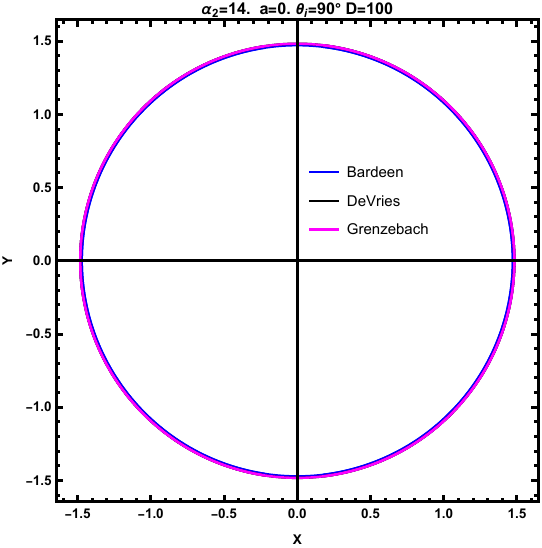}
        \caption{\label{geometric_alpha_14._a_0._plasma_inhomogeneous_D_100_theta_90.pdf}}
    \end{subfigure}
    \hfill
    \begin{subfigure}{0.25\textwidth}
    \includegraphics[width=\linewidth,height=\textwidth]{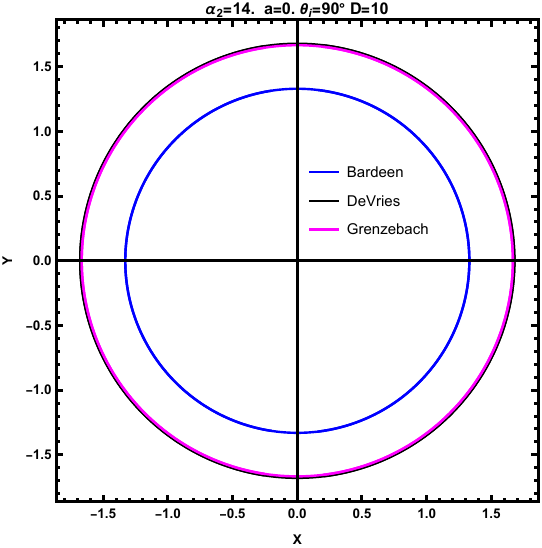}
        \caption{\label{geometric_alpha_14._a_0._plasma_inhomogeneous_D_10_theta_90.pdf}}
    \end{subfigure}
    \hfill
    \begin{subfigure}{0.25\textwidth}
    \includegraphics[width=\linewidth,height=\textwidth]{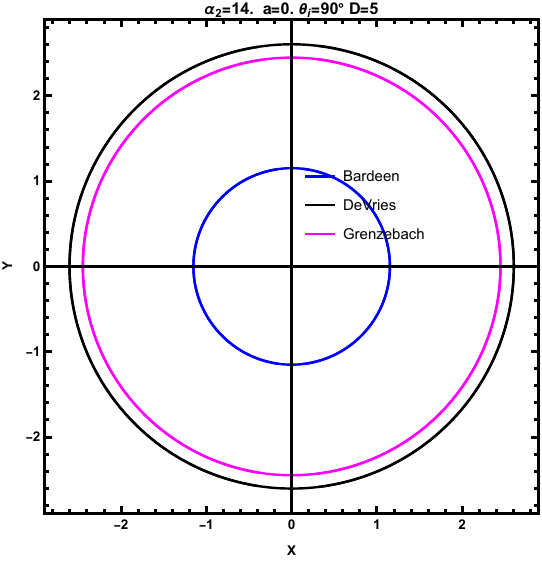}
        \caption{\label{geometric_alpha_14._a_0._plasma_inhomogeneous_D_5_theta_90.pdf}}
    \end{subfigure}  
    \hfill
    \begin{subfigure}{0.25\textwidth}
    \includegraphics[width=\linewidth,height=\textwidth]{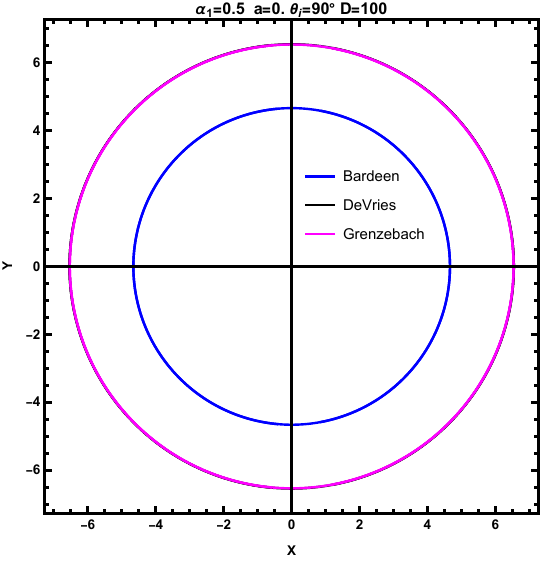}
        \caption{\label{geometric_alpha_0.5_a_0._plasma_homogeneous_D_100_theta_90.pdf}}
    \end{subfigure}
    \hfill
    \begin{subfigure}{0.25\textwidth}
    \includegraphics[width=\linewidth,height=\textwidth]{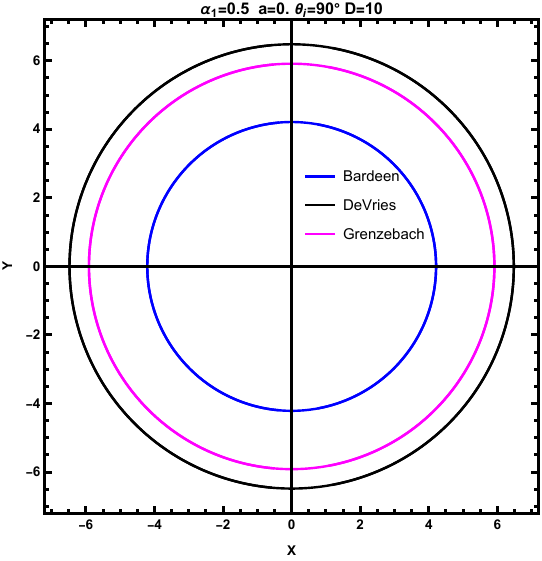}
        \caption{\label{geometric_alpha_0.5_a_0._plasma_homogeneous_D_10_theta_90.pdf}}
    \end{subfigure}
    \hfill
    \begin{subfigure}{0.25\textwidth}
    \includegraphics[width=\linewidth,height=\textwidth]{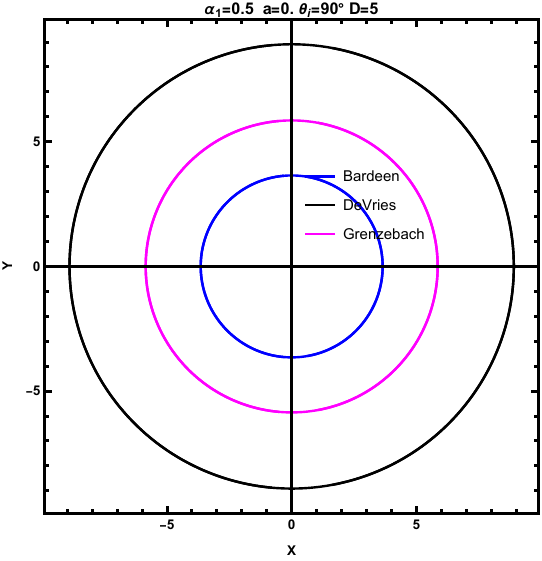}
        \caption{\label{geometric_alpha_0.5_a_0._plasma_homogeneous_D_5_theta_90.pdf}}
    \end{subfigure}   
       
    \caption{Above figure shows the critical curve of a Schwarzschild black hole   computed using Bardeen's definition  $(X_B,Y_B)$ (blue color), de Vries  definition $(X_D,Y_D)$ (black color) and Grenzebach's definition $(X_G,Y_G)$ (magenta color). Each row shows the variation of the critical curve   with distance (decreasing from left to right). The first row corresponds to the vacuum case, the second row corresponds to the case of inhomogeneous plasma and the third row represents the case of homogeneous plasma. \label{Schwarzschild plasma} }
\end{figure}

%%%%%%%%%%%%%%%%%%%%%%%%%%%%%%%%%%%%%%%%%%%%%%%%%%%%
\begin{figure}[t!]
    \begin{subfigure}{0.25\textwidth}
    \includegraphics[width=\linewidth,height=\textwidth]{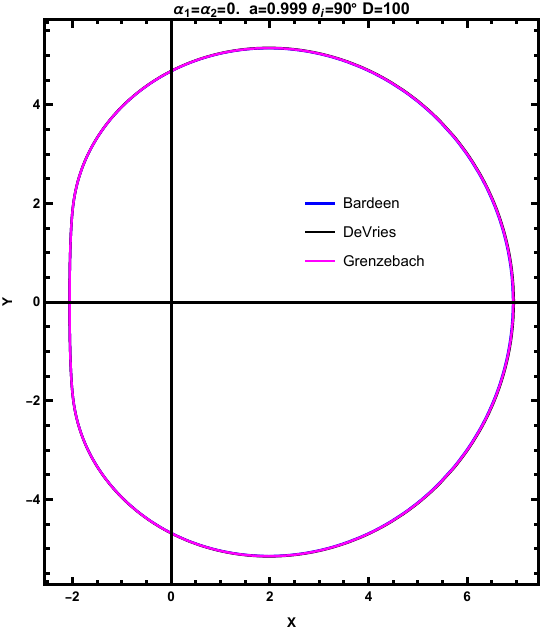}
        \caption{\label{geometric_alpha_0._a_0.999_plasma_homogeneous_D_100_theta_90.pdf}}
    \end{subfigure}
    \hfill
    \begin{subfigure}{0.25\textwidth}
    \includegraphics[width=\linewidth,height=\textwidth]{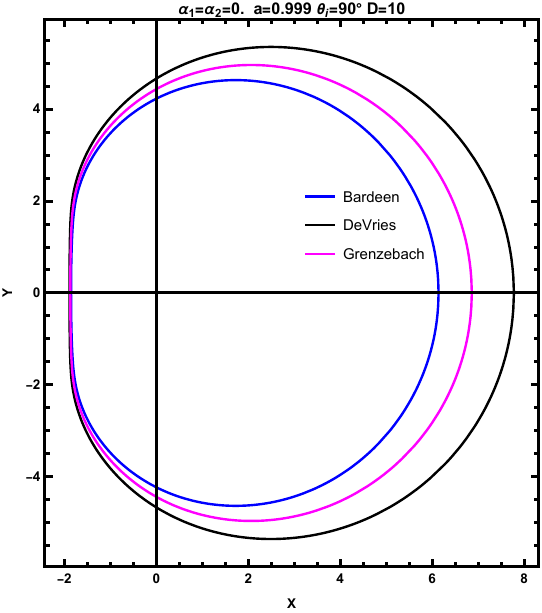}
        \caption{\label{geometric_alpha_0._a_0.999_plasma_homogeneous_D_10_theta_90.pdf}}
    \end{subfigure}
    \hfill
    \begin{subfigure}{0.25\textwidth}
    \includegraphics[width=\linewidth,height=\textwidth]{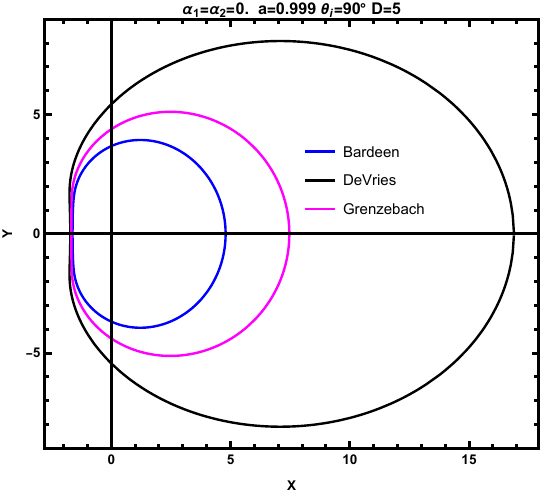}
        \caption{\label{geometric_alpha_0._a_0.999_plasma_homogeneous_D_5_theta_90.pdf}}
    \end{subfigure}
     \hfill
    \begin{subfigure}{0.25\textwidth}
    \includegraphics[width=\linewidth,height=\textwidth]{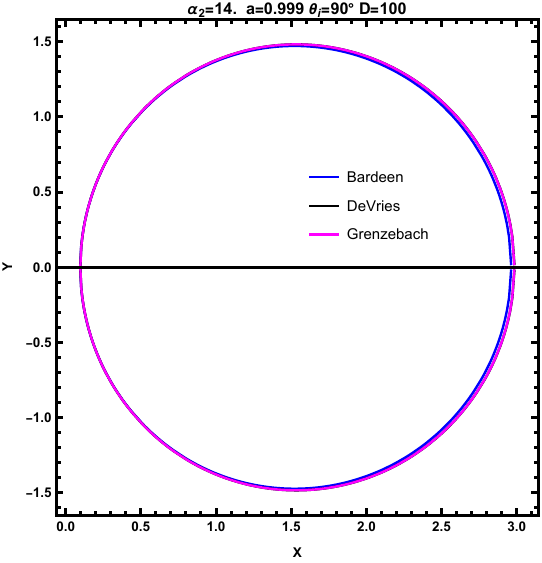}
        \caption{\label{geometric_alpha_14._a_0.999_plasma_inhomogeneous_D_100_theta_90.pdf}}
    \end{subfigure}
    \hfill
    \begin{subfigure}{0.25\textwidth}
    \includegraphics[width=\linewidth,height=\textwidth]{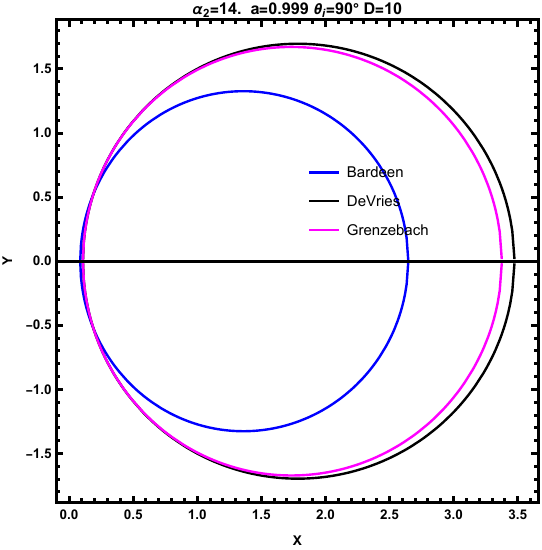}
        \caption{\label{geometric_alpha_14._a_0.999_plasma_inhomogeneous_D_10_theta_90.pdf}}
    \end{subfigure}
    \hfill
    \begin{subfigure}{0.25\textwidth}
    \includegraphics[width=\linewidth,height=\textwidth]{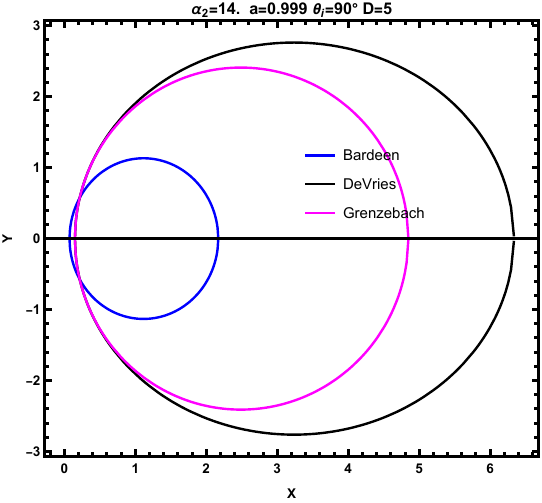}
        \caption{\label{geometric_alpha_14._a_0.999_plasma_inhomogeneous_D_5_theta_90.pdf}}
    \end{subfigure} 
    \hfill
    \begin{subfigure}{0.25\textwidth}
    \includegraphics[width=\linewidth,height=\textwidth]{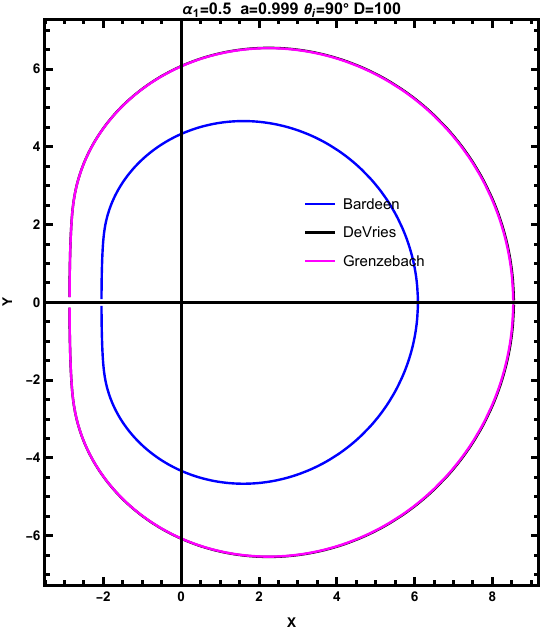}
        \caption{\label{geometric_alpha_0.5_a_0.999_plasma_homogeneous_D_100_theta_90.pdf}}
    \end{subfigure}
    \hfill
    \begin{subfigure}{0.25\textwidth}
    \includegraphics[width=\linewidth,height=\textwidth]{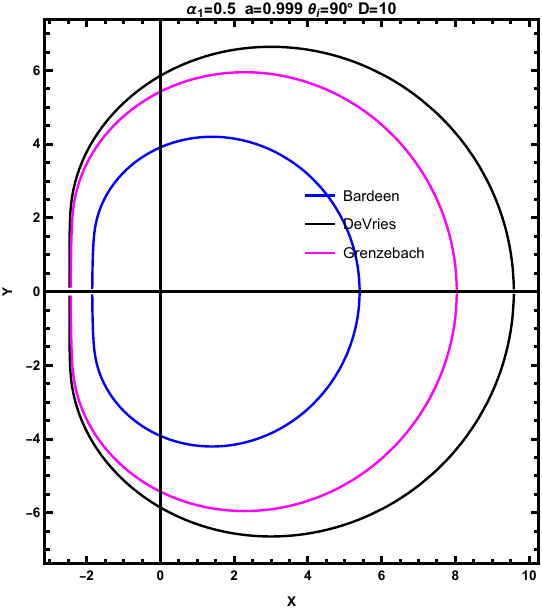}
        \caption{\label{geometric_alpha_0.5_a_0.999_plasma_homogeneous_D_10_theta_90.pdf}}
    \end{subfigure}
    \hfill
    \begin{subfigure}{0.25\textwidth}
    \includegraphics[width=\linewidth,height=\textwidth]{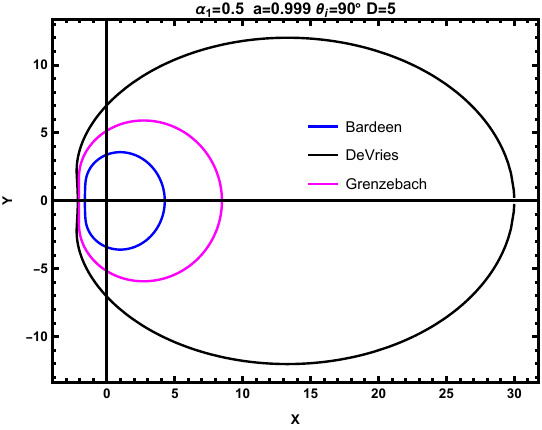}
        \caption{\label{geometric_alpha_0.5_a_0.999_plasma_homogeneous_D_5_theta_90.pdf}}
    \end{subfigure}   
       
    \caption{Above figure shows the critical curve of a near extremal Kerr black hole   computed using Bardeen's definition  $(X_B,Y_B)$ (blue color), de Vries  definition $(X_D,Y_D)$ (black color) and Grenzebach's definition $(X_G,Y_G)$ (magenta color). Each row shows the variation of the critical curve   with distance (decreasing from left to right). The first row corresponds to the vacuum case, the second row corresponds to the case of inhomogeneous plasma and the third row represents the case of homogeneous plasma.  Here we take observer inclination angle to be $\theta_i=90^\circ$. \label{Kerr plasma} }
\end{figure}

\begin{itemize}
    \item As we go from left to right in the first row  (\cref{geometric_alpha_0._a_0._plasma_homogeneous_D_100_theta_90.pdf,geometric_alpha_0._a_0._plasma_homogeneous_D_10_theta_90.pdf,geometric_alpha_0._a_0._plasma_homogeneous_D_5_theta_90.pdf,geometric_alpha_0._a_0.999_plasma_homogeneous_D_100_theta_90.pdf,geometric_alpha_0._a_0.999_plasma_homogeneous_D_10_theta_90.pdf,geometric_alpha_0._a_0.999_plasma_homogeneous_D_5_theta_90.pdf}) and  the second row (\cref{geometric_alpha_14._a_0._plasma_inhomogeneous_D_100_theta_90.pdf,geometric_alpha_14._a_0._plasma_inhomogeneous_D_10_theta_90.pdf,geometric_alpha_14._a_0._plasma_inhomogeneous_D_5_theta_90.pdf,geometric_alpha_14._a_0.999_plasma_inhomogeneous_D_100_theta_90.pdf,geometric_alpha_14._a_0.999_plasma_inhomogeneous_D_10_theta_90.pdf,geometric_alpha_14._a_0.999_plasma_inhomogeneous_D_5_theta_90.pdf}) of \cref{Schwarzschild plasma,Kerr plasma},  we observe that the difference between the critical curves computed using $(X_B,Y_B)$ in \cref{Bardeen tetrad XB,Bardeen tetrad YB}, $(X_D,Y_D)$ in \cref{Bardeen tetrad XD,Bardeen tetrad YD} and $(X_G,Y_G)$ in \cref{Bardeen tetrad XG,Bardeen tetrad YG} increases as the observer distance decreases. 

    \item  In all individual plots for $D=10 \text{ and  5}$  of \cref{Schwarzschild plasma,Kerr plasma}, when the observer distance is comparable to critical curve size, we observe $(X_D,Y_D)>(X_G,Y_G)>(X_B,Y_B)$  as predicted from analysis in previous section (refer discussion after \cref{finite XB gamma delta,finite YB gamma delta}).  The main difference between the plots in  \cref{Schwarzschild plasma,Kerr plasma} is that the critical curves in \cref{Kerr plasma} are distorted compared to that in \cref{Schwarzschild plasma} because of the combined effect of  the spin of the black hole and  the inclination of the observer. 
    \item   When $D=100$ we observe that  $(X_B,Y_B)$, $(X_D,Y_D )$ and $(X_G,Y_G)$ nearly coincide in the vacuum case and in the case of inhomogeneous plasma.  However, when homogeneous plasma is considered, only $(X_D,Y_D)$ and $(X_G,Y_G)$ nearly coincide, but $(X_B,Y_B)$ is much smaller than the critical curve in the vacuum case. This mismatch is expected to hold even at very large distances. Thus, homogeneous plasma clearly highlights the inconsistency between $(X_B,Y_B)$ with  $(X_D,Y_D)$ and $(X_G,Y_G)$ and also studies from gravitational lensing. This has been discussed in detail in the previous section (refer to the discussion after \cref{large D XG,large D YG}).

    \item   As we go from top to bottom in each column of \cref{Schwarzschild plasma,Kerr plasma}, we observe that inhomogeneous plasma produces a contracted critical curve, and homogeneous plasma produces a magnified critical curve in comparison to  the vacuum case when we use de Vries and Grenzebach's definition of the critical curve   $\big((X_D,Y_D)$ and $(X_G,Y_G)\big)$. In contrast to this, when Bardeen's definition  $(X_B,Y_B)$ is used, we get a contracted critical curve irrespective of plasma profile being  inhomogeneous or homogeneous.

\end{itemize}

The inconsistencies between Bardeen's definition $(X_B,Y_B)$, de Vries's definition $(X_D,Y_D)$ and Grenzebach's definition $(X_G,Y_G)$  observed in  \cref{Kerr plasma,Schwarzschild plasma} had already been highlighted and discussed in  \cref{Comparison of different methods section}.      In fact, in \cref{Comparison of different methods section}  our main assumptions were, the black hole is Kerr-like, the observer is stationary, and the tetrad used to describe the  observer's frame is stationary, orthonormal, and of the form \citep{1993GReGrO,Bardeen:1972fi,Carter:1968rr}   \cref{stationary tetrads et,stationary tetrads er,stationary tetrads e theta,stationary tetrads e phi}. Thus, the discussion in \cref{Comparison of different methods section} is not only limited to Bardeen tetrads but also applies to other stationary orthonormal tetrads \citep{1993GReGrO}.   Additionally, we deduced and discussed the  inconsistencies   without invoking the separability condition  \cref{separable plasma condition}  throughout our analysis  from    \cref{tetrad expansion DeVreis,tetrad expansion grenzebach,alpha,beta,gamma,delta,alpha beta omega,tan x,tan y,finite XB gamma delta,finite YB gamma delta,finite XD in gamma delta,finite YD in gamma delta,large D XB,large D YB,large D XD,large D YD,large D XG,large D YG}.  Finally, we were also able to deduce  de Vries's definition $(X_D,Y_D)$ by assuming a critical curve \citep{Gralla:2019xty,Perlick:2021aok} and thus provide an explanation for the possible discrepancy between de Vries's definition $(X_D,Y_D)$ and Grenzebach's definition $(X_G,Y_G)$. Hence, the main objective of this section was to visualise these inconsistencies for the case of the Schwarzschild and  Kerr black holes; in particular, to highlight that the inconsistency arising due to finite distance is significant when the observer distance $D$ is of the order  of the size of the critical curve. More importantly, the contradictory nature of the effect of  homogeneous plasma on the critical curve size, as predicted by Bardeen's definition, persists even when the  distance increases.   

\section{Analysing the critical curve of black holes using Carter tetrad \label{Analysing the critical curve using Carter tetrad}}
In  this section, we will compare the critical curves of a Kerr black hole  surrounded by plasma profiles 1 and 2, computed in Carter tetrads \citep{Carter:1968rr,1993GReGrO}  with those computed in Bardeen tetrads \citep{Bardeen:1972fi,1977MNRASZnajek,1993GReGrO} (\cref{Analysing the critical curve using Bardeen tetrad}). We will first obtain expressions of $(X_B,Y_B)$, $(X_b,Y_b)$,  $(X_D,Y_D)$, $(X_d,Y_d)$, $(X_G,Y_G)$ and $(X_g,Y_g)$ in the Carter frame.  The Carter tetrad for an observer at a distance $D$ and inclination $\theta_i$ is given by:
\begin{align}
    \label{carter tetrad et}
    e_{(t)}&=\left( \frac{ \Sigma}{\rho\sqrt{\Delta}},0,0,\frac{ a}{\rho\sqrt{\Delta}}\right)\Bigg|_{(r=D,\theta=\theta_i)}\\
    \label{carter tetrad er}
    e_{(r)}&=\left(0,\frac{\sqrt{\Delta}}{\rho},0,0\right)\Bigg|_{(r=D,\theta=\theta_i)}\\
    \label{carter tetrad e theta}
     e_{(\theta)}&=\left(0,\frac{1}{\rho},0,0\right)\Bigg|_{(r=D,\theta=\theta_i)}\\
    \label{carter tetrad e phi}
     e_{(\phi)}&=\left( \frac{ a\sin\theta}{\rho},0,0,\frac{ 1}{\rho\sin{\theta}}\right)\Bigg|_{(r=D,\theta=\theta_i)}
\end{align}
Using the  above tetrads and proceeding similarly to the previous section, the tetrad components of momenta $p^{(a)}$ and the celestial angles, $\gamma$ and $\delta$  for a Carter observer are given by:
\begin{align}
    \label{Carter pt}
    \frac{p^{(t)}}{E}&=\frac{\Sigma-a\eta}{\sqrt{\rho^2\Delta}}\Bigg|_{(r=D,\theta=\theta_i)}\\
    \label{Carter pr}
    \frac{p^{(r)}}{E}&=\sqrt{\frac{(\Sigma-a\eta)^2-(\chi+f(r))\Delta}{\rho^2\Delta}}\Bigg|_{(r=D,\theta=\theta_i)}\\
    \label{Carter p theta}
    \frac{p^{(\theta)}}{E}&=\sqrt{\frac{1}{\rho^2}\left(\chi-g(\theta)-\left(\frac{\eta}{\sin\theta}-a\sin\theta\right)^2\right)}\Bigg|_{(r=D,\theta=\theta_i)}\\
    \label{Carter p phi}
    \frac{p^{(\phi)}}{E}&=\frac{1}{\rho}\left(\frac{\eta}{\sin\theta}-a\sin\theta\right)\Bigg|_{(r=D,\theta=\theta_i)}\\
    \label{Carter gamma}
    \gamma&=\arctan{\left(\sqrt{\frac{\Delta}{V(r)}\left(\chi-g(\theta)\right)}\right)}\Bigg|_{(r=D,\theta=\theta_i)}\\
    \label{Carter delta}
    \delta&=\arctan{\left(\frac{-\sqrt{\left(\chi-g(\theta)-\left({\eta\csc\theta}-a\sin\theta\right)^2\right)}}{(\eta\csc\theta-a\sin\theta)}\right)}\Bigg|_{(r=D,\theta=\theta_i)}
\end{align}
For a finite distance observer, the expressions of $(X_B,Y_B)$, $(X_D,Y_D)$ and $(X_G,Y_G)$ are given by:
\begin{align}
    \label{Carter tetrad XB}
  X_B&=\frac{-r\sqrt{\Delta}\ (\eta\csc\theta-a\sin\theta)}{\Sigma-a\eta}\Bigg|_{(r=D,\theta=\theta_i)}\\
  \label{Carter tetrad YB}
  Y_B&=\frac{r\sqrt{\Delta W(\theta)}}{\Sigma-a\eta}\Bigg|_{(r=D,\theta=\theta_i)}\\
  \label{Carter tetrad XD}
  X_D&=\frac{-r\sqrt{\Delta}\ (\eta\csc\theta-a\sin\theta)}{\sqrt{V(r)}}\Bigg|_{(r=D,\theta=\theta_i)}\\
  \label{Carter tetrad YD}
  Y_D&=\frac{r\sqrt{\Delta\left(\chi-g(\theta)-(\eta\csc\theta-a\sin\theta)^2\right)}}{\sqrt{V(r)}}\Bigg|_{(r=D,\theta=\theta_i)}\\
   \label{Carter tetrad XG}
    X_G&=D\ \tilde X_G=2 D \tan\left(\frac{\gamma}{2}\right) \sin\delta\\
    \label{Carter tetrad YG}
    Y_G&=D\ \tilde Y_G= 2 D \tan\left(\frac{\gamma}{2}\right) \cos\delta
\end{align}
    where, $\gamma$ and $\delta$ are given by \cref{Carter gamma,Carter delta}. For an observer located at infinity (or $D$ is large), the parametric equations of the  critical curves  computed using $(X_B,Y_B)$, $(X_D,Y_D)$ and $(X_G,Y_G)$ are given by
 \begin{align}
     \label{Carter Xb}
     X_b&=\lim_{D\rightarrow\infty}X_B=-\left(\frac{\eta}{\sin\theta_i}-a\sin\theta_i\right)\\
     \label{CarterYb}
     Y_b&=\lim_{D\rightarrow\infty}Y_B=\sqrt{\chi-g(\theta_i)-\left(\eta\csc\theta_i-a\sin\theta_i\right)^2}\\
     \label{Carter Xd}
     X_d&=\lim_{D\rightarrow\infty}X_D=\frac{-(\eta\csc\theta_i-a\sin\theta_i)}{\sqrt{1-\lambda}}\\
     \label{Carter Yd}
     Y_d&=\lim_{D\rightarrow\infty}Y_D=\frac{\sqrt{\chi-g(\theta_i)-\left(\eta\csc\theta_i-a\sin\theta_i\right)^2}}{\sqrt{1-\lambda}}\\
     &\text{where, }\lambda=\lim_{D\rightarrow\infty}\frac{f(r)}{r^2}\Bigg|_{r=D}\nonumber\\
     \label{Carter Xg}
     X_g&=\lim_{D\rightarrow\infty}X_G=X_d\\
     \label{Carter Yg}
     Y_g&=\lim_{D\rightarrow\infty}Y_G=Y_d      
     \end{align}

The horizontal shift in the critical curve due to the change in tetrad can be seen  by comparing \cref{Carter Xb,Carter Xd} with \cref{Xb,Xd}. Notice that  \cref{Carter gamma}  and   \cref{Bardeen gamma}  reduce to the same expression  for the case of a Schwarzschild black hole. This is because, in the limit $a\rightarrow0$, both Carter and Bardeen  tetrads reduce to the  static tetrad \citep{1993GReGrO}. This is also the reason for the lack of horizontal shift of the  critical curve and the  origin of the observer's sky plane for the case of a Schwarzschild black hole, irrespective of shifting from Bardeen to Carter tetrads. Additionally, the celestial angle $\gamma$ also becomes the angular radius of the critical curve in the case of a Schwarzschild black hole (refer \cref{gamma _and_delta.png}). 

The  angular radius $\gamma _{S}$  of the critical curve of a  Schwarzschild black hole surrounded by a plasma satisfying the separability condition \cref{separable plasma condition} (with $a=0$) is given by   \citep{Perlick:2021aok};
\begin{align}
    \label{angular radius Schwarzschild}
    \gamma_{S}&=\arctan\left(\sqrt{\frac{(r-2 )(\chi_S-g(\theta))}{r^3-(\chi_S+f(r))\ (r-2)}}\right)\Bigg|_{(r=D,\theta=\theta_i)}\\
    \text{where,  }\chi_S&=\frac{r_{sp}^3}{r_{sp}-2 }-f(r_{sp})\text{ \citep{Perlick:2017fio} and $r_{sp}$ obtained by solving $a\eta(r_{sp})=0$ \citep{Perlick:2017fio,KumarSahoo:2025igt} in \cref{kerr a eta}.}\nonumber
\end{align}
For an observer at a finite distance, the geometric radius of the critical curve $R_S$ is given by:
\begin{align}
    \label{geometric radius Schwarzschild plasma finite distance}
    R_S&=D\tan\gamma_S=D\sqrt{\frac{(D-2)(\chi_S-g(\theta_i))}{D^3-(\chi_S+f(D))\ (D-2)}}
\end{align}
when the  distance becomes large, i. e.,  $D\rightarrow\infty$, then the  geometric radius $R_{S\infty}$ is given by:
\begin{align}
 \label{geometric radius Schwarzschild plasma infinite distance}
    R_{S\infty}&=\lim_{D\rightarrow\infty}R_S=\sqrt{\frac{\chi_S-g(\theta_i)}{1-\lambda}}\\
    \text{where, }\lambda&=\lim_{D\rightarrow\infty}\frac{f(D)}{D^2}\nonumber
\end{align}
when plasma is absent  ($f(r)=0\text{, }g(\theta)=0$), $r_{sp}=3\text{ and }\chi_S=27$ \citep{Perlick:2015vta,Perlick:2017fio}. Thus, in the absence of plasma   \cref{angular radius Schwarzschild,geometric radius Schwarzschild plasma finite distance,geometric radius Schwarzschild plasma infinite distance} reduces to:
\begin{align}
    \label{angular radius Schwarzschild no plasma}
    \tan\gamma_S&=\sqrt{ \frac{{\left(1-\frac{2}{D}\right)}}{{\left(\frac{2}{D}-1+\frac{D^2}{27}\right)}}}\\
     \label{geometric radius Schwarzschild no plasma finite distance}
    R_S&=D\tan\gamma_S=D\sqrt{\frac{27 (D-2)}{D^3-27\ D+54}}\\
     \label{geometric radius Schwarzschild no plasma infinite distance}
    R_{S\infty}&=\lim_{D\rightarrow\infty}R_S=3\sqrt{3}
\end{align}
The expression   \cref{angular radius Schwarzschild no plasma} was  obtained by Zel'dovich and Novikov \citep{Zeldovich1966} and is equivalent to Synge's formula \citep{Synge:1966okc,Perlick:2021aok}.   

The geometric radius $R_S$ of a circular critical curve also represents the distance of a point on the critical curve from the origin of the observer's sky plane. Thus, using \cref{finite XD in gamma delta,finite YD in gamma delta} it can be shown that $R_S$ is  directly related  only to  $(X_D,Y_D)$  at a finite distance as;
\begin{align}
    \label{Rs XD YD}
    R_S=\sqrt{X^2_D+Y^2_D}=D\ \tan\gamma_S
\end{align}
and not to $(X_B,Y_B)$ and $(X_G,Y_G)$ because,
\begin{align}
    \label{RS not XB YB}
    \sqrt{X^2_B+Y^2_B}&=\left(\sqrt{1-\frac{\omega^2_P}{\alpha^2}}\right)D\ \sin\gamma_S\neq R_S\\
\label{RS not XG YG}
    \sqrt{X^2_G+Y^2_G}&=2D\ \tan\left(\frac{\gamma_S}{2}\right)\neq R_S
\end{align}
However, at a large distance $D$ and when homogeneous plasma is not considered $\sqrt{X^2_B+Y^2_B}\approx\sqrt{X^2_G+Y^2_G}\approx R_S$. 

We now proceed to analyse the critical curve in the observer's sky plane computed using Bardeen's definition \citep{Bardeen:1973}, de Vries's definition \citep{2025CQGHioki} and Grenzebach et al.'s definition \citep{Grenzebach:2014fha,Grenzebach2015book} in Carter tetrads. Since the inconsistencies between $(X_B,Y_B)$, $(X_D,Y_D)$ and $(X_G,Y_G)$ are generic ( as discussed in \cref{Comparison of different methods section}), we will now focus on distances near the horizon and distances at an astrophysical scale. For this, we consider  cases  when the equatorial observer is at $D=5$, and $D\approx5.64\times10^{10}$, which is computed using the distance of earth from M87*,  $d=16.8$Mpc  and its  mass $M=6.2\times10^9M_{\odot}$ \citep{EventHorizonTelescope:2019dse}.
In \cref{Schwarzschild plasma carter,Kerr plasma carter}, plots of critical curves computed  using $(X_B,Y_B)$, $(X_D,Y_D)$ and $(X_G,Y_G)$ with Carter and Bardeen tetrads are shown  for Schwarzschild and extremal Kerr black holes, respectively. The critical curves with Carter  tetrads are solid curves, and the critical curves with Bardeen tetrads are dashed curves. As was done in  the previous section, the critical curves computed using $(X_B,Y_B)$, $(X_D,Y_D)$ and $(X_G,Y_G)$ are coloured blue, black, and magenta, respectively.

As discussed in the beginning of this section and also evident from \cref{Schwarzschild plasma carter}, the critical curves of a Schwarzschild black hole computed in Carter and Bardeen tetrads completely coincide. As expected, the inconsistencies between $(X_B,Y_B)$, $(X_D,Y_D)$ and $(X_G,Y_G)$ are clearly visible in \cref{carter_geometric_alpha_0._a_0._plasma_homogeneous_D_5_theta_90.pdf,carter_geometric_alpha_14._a_0._plasma_inhomogeneous_D_5_theta_90.pdf,carter_geometric_alpha_0.5_a_0._plasma_homogeneous_D_5_theta_90.pdf} when $D=5$.   For the near horizon observer, $(X_B,Y_B)$, $(X_D,Y_D)$ and $(X_G,Y_G)$ generate critical curves of different sizes, with $(X_D,Y_D)$ being the largest and $(X_B,Y_B)$ being the smallest (similar to the Bardeen tetrads shown in the previous section).  Additionally,   \cref{carter_geometric_alpha_0._a_0._plasma_homogeneous_D_m87_theta_90.pdf,carter_geometric_alpha_14._a_0._plasma_inhomogeneous_D_m87_theta_90.pdf,carter_geometric_alpha_0.5_a_0._plasma_homogeneous_D_m87_theta_90.pdf}   represent critical curves for an equatorial observer at a distance $d=16.8$Mpc  from M87*. We find that even for a distant observer, $(X_B,Y_B)$ produces a smaller critical curve when  homogeneous plasma  is considered  as compared to the case of no plasma, in contradiction with $(X_D,Y_D)$ and $(X_G,Y_G)$ and also previously reported studies. As deduced  in \cref{Comparison of different methods section}, it can be observed in    \cref{carter_geometric_alpha_0._a_0._plasma_homogeneous_D_m87_theta_90.pdf,carter_geometric_alpha_14._a_0._plasma_inhomogeneous_D_m87_theta_90.pdf} that for a distant observer, when inhomogeneous plasma is considered $(X_B,Y_B)$, $(X_G,Y_G)$ and $(X_D,Y_D)$ tend to give the same critical curve. The only additional feature as compared to \cref{Schwarzschild plasma carter}  observed in \cref{Kerr plasma carter}, is the horizontal shift between the critical curves computed using  Bardeen (dashed curves)  and Carter   tetrads  (solid curves) due to the combined effect of the spin $a$ and inclination $\theta_i$.

\begin{figure}[H]
    \begin{subfigure}{0.5\textwidth}
    \centering
    \includegraphics[scale=0.5]{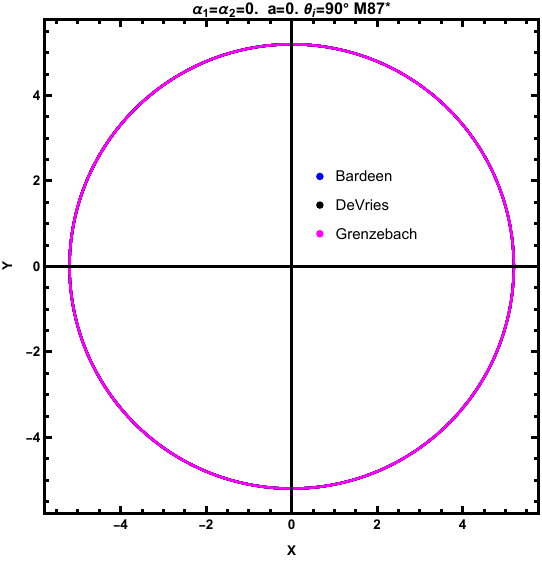}
        \caption{\label{carter_geometric_alpha_0._a_0._plasma_homogeneous_D_m87_theta_90.pdf}}
    \end{subfigure}
    \hfill
    \begin{subfigure}{0.5\textwidth}
    \centering
    \includegraphics[scale=0.5]{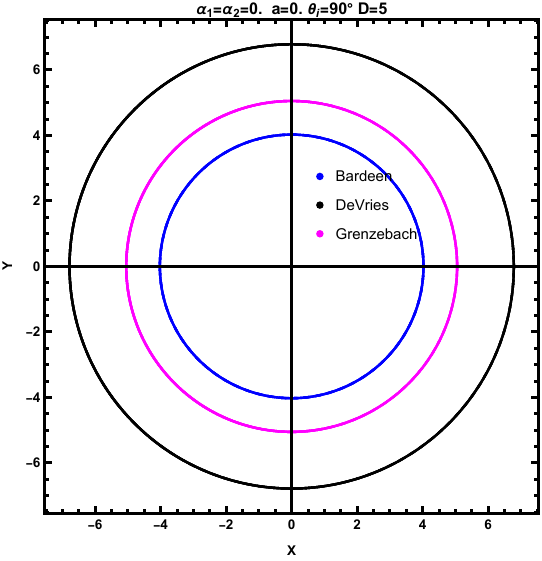}
        \caption{\label{carter_geometric_alpha_0._a_0._plasma_homogeneous_D_5_theta_90.pdf}}
    \end{subfigure}
     \hfill
    \begin{subfigure}{0.5\textwidth}
    \centering
    \includegraphics[scale=0.5]{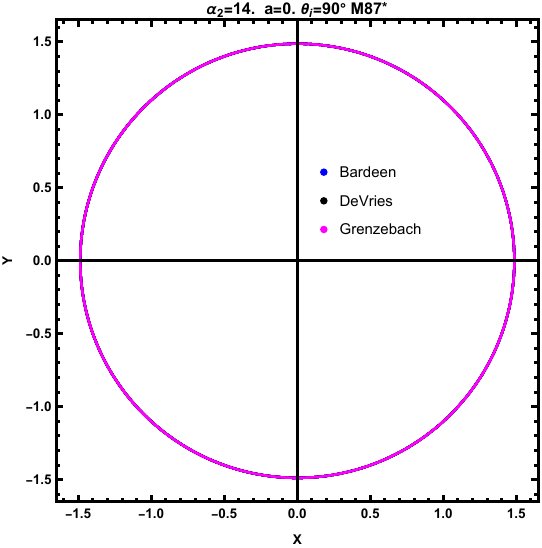}
        \caption{\label{carter_geometric_alpha_14._a_0._plasma_inhomogeneous_D_m87_theta_90.pdf}}
    \end{subfigure}
    \hfill
    \begin{subfigure}{0.5\textwidth}
    \centering
    \includegraphics[scale=0.5]{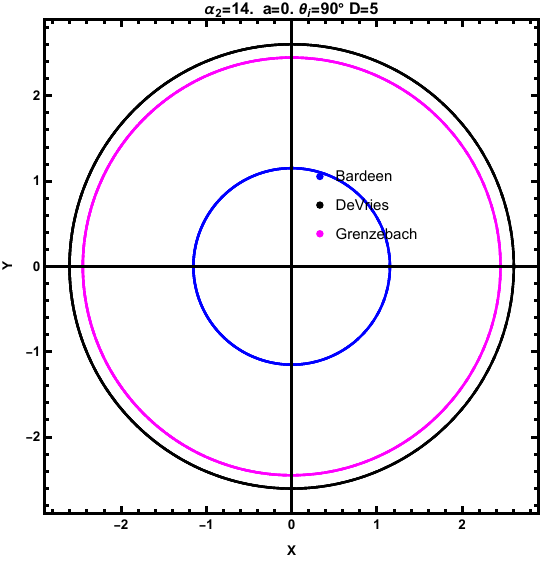}
        \caption{\label{carter_geometric_alpha_14._a_0._plasma_inhomogeneous_D_5_theta_90.pdf}}
    \end{subfigure} 
    \hfill
    \begin{subfigure}{0.5\textwidth}
    \centering
    \includegraphics[scale=0.5]{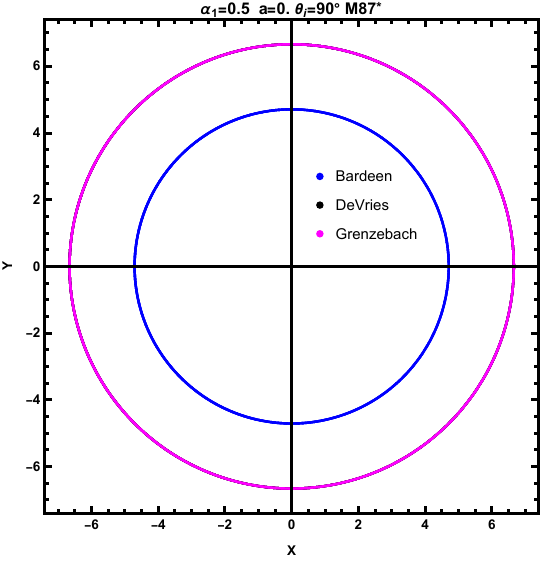}
        \caption{\label{carter_geometric_alpha_0.5_a_0._plasma_homogeneous_D_m87_theta_90.pdf}}
    \end{subfigure}
    \hfill
    \begin{subfigure}{0.5\textwidth}
    \centering
    \includegraphics[scale=0.5]{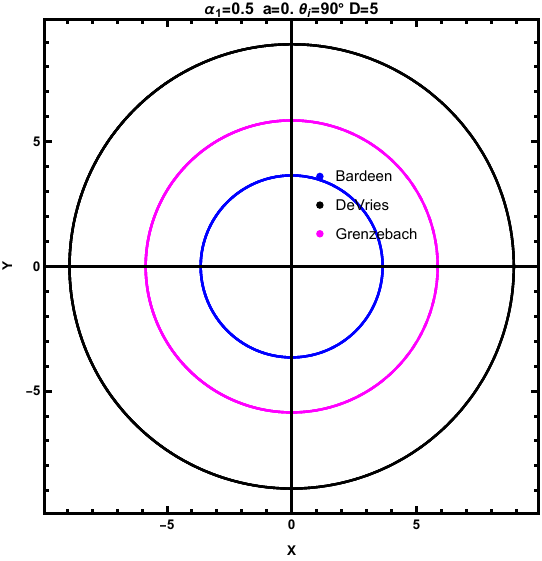}
        \caption{\label{carter_geometric_alpha_0.5_a_0._plasma_homogeneous_D_5_theta_90.pdf}}
    \end{subfigure} 
    \caption{Above figure shows the critical curve of a Schwarzschild black hole   computed using Bardeen's definition  $(X_B,Y_B)$ (blue color), de Vries  definition $(X_D,Y_D)$ (black color) and Grenzebach's definition $(X_G,Y_G)$ (magenta color). Each row shows the variation of the critical curve   with distance (decreasing from left to right). The first row corresponds to the vacuum case, the second row corresponds to the case of inhomogeneous plasma and the third row represents the case of homogeneous plasma. The dashed curves  computed using Bardeen tetrads and the solid curves  computed using Carter tetrads  coincide. Both dashed and solid curves are indistinguishable because, for the Schwarzschild black hole the Bardeen and Carter tetrad reduce to the static tetrad.  \label{Schwarzschild plasma carter} }
\end{figure}
\begin{figure}[H]
    \begin{subfigure}{0.5\textwidth}
    \centering
    \includegraphics[scale=0.5]{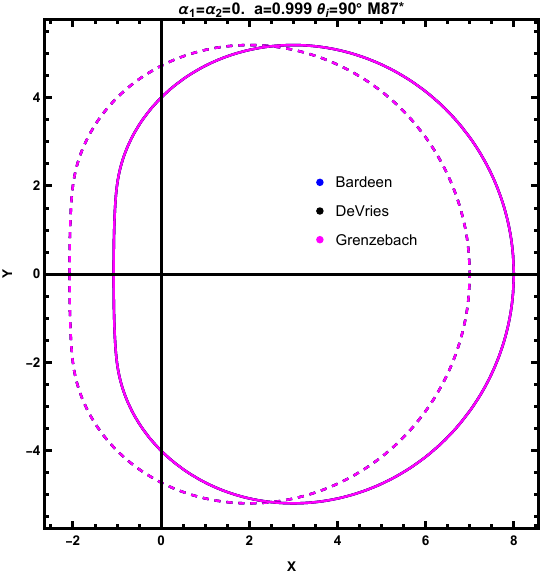}
        \caption{\label{carter_geometric_alpha_0._a_0.999_plasma_homogeneous_D_m87_theta_90.pdf}}
    \end{subfigure}
    \hfill
    \begin{subfigure}{0.5\textwidth}
    \centering
    \includegraphics[scale=0.5]{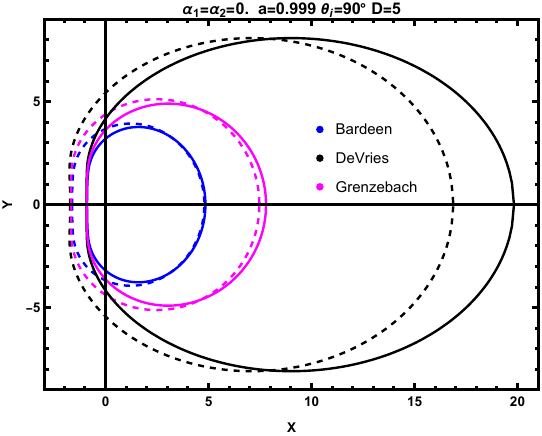}
        \caption{\label{carter_geometric_alpha_0._a_0.999_plasma_homogeneous_D_5_theta_90.pdf}}
    \end{subfigure}
     \hfill
    \begin{subfigure}{0.5\textwidth}
    \centering
    \includegraphics[scale=0.5]{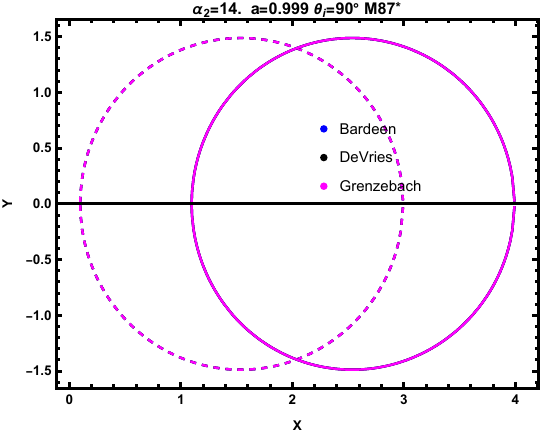}
        \caption{\label{carter_geometric_alpha_14._a_0.999_plasma_inhomogeneous_D_m87_theta_90.pdf}}
    \end{subfigure}
    \hfill
    \begin{subfigure}{0.5\textwidth}
    \centering
    \includegraphics[scale=0.5]{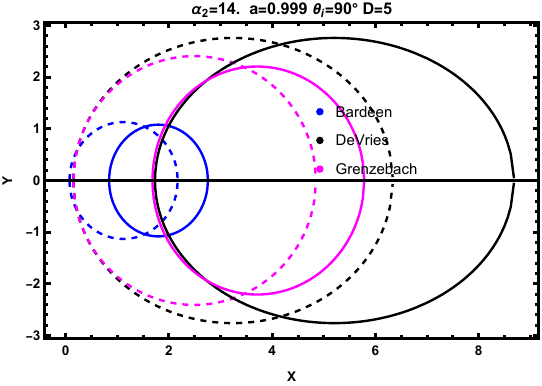}
        \caption{\label{carter_geometric_alpha_14._a_0.999_plasma_inhomogeneous_D_5_theta_90.pdf}}
    \end{subfigure} 
    \hfill
    \begin{subfigure}{0.5\textwidth}
    \centering
    \includegraphics[scale=0.5]{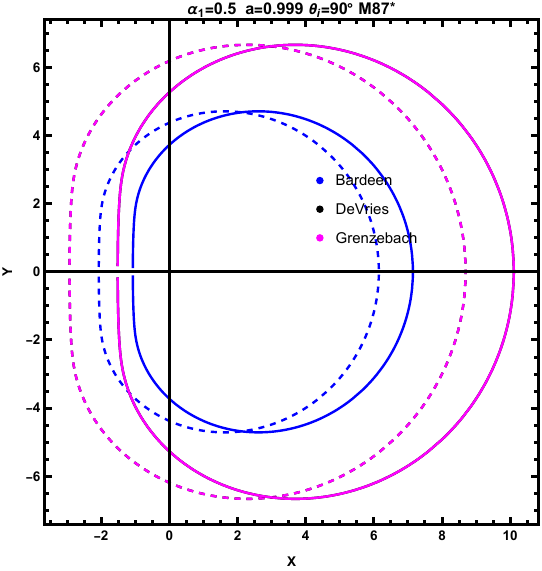}
        \caption{\label{carter_geometric_alpha_0.5_a_0.999_plasma_homogeneous_D_m87_theta_90.pdf}}
    \end{subfigure}
    \hfill
    \begin{subfigure}{0.5\textwidth}
    \centering
    \includegraphics[scale=0.5]{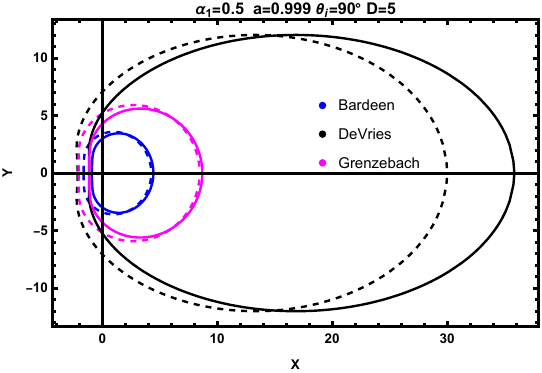}
        \caption{\label{carter_geometric_alpha_0.5_a_0.999_plasma_homogeneous_D_5_theta_90.pdf}}
    \end{subfigure} 
    \caption{Above figure shows the critical curve of a near extremal Kerr black hole   computed using Bardeen's definition  $(X_B,Y_B)$ (blue color), de Vries  definition $(X_D,Y_D)$ (black color) and Grenzebach's definition $(X_G,Y_G)$ (magenta color). Each row shows the variation of the critical curve   with distance (decreasing from left to right). The first row corresponds to the vacuum case, the second row corresponds to the case of inhomogeneous plasma and the third row represents the case of homogeneous plasma. The dashed curves have been computed using Bardeen tetrads and the solid curves have been computed using Carter tetrads.  Here the observer inclination angle is taken to be $\theta_i=90^\circ$.}\label{Kerr plasma carter} 
\end{figure}

\section{Conclusion\label{Conclusion}}

In the present work,  we   compared the definitions  by Bardeen \citep{Bardeen:1973,Chandrasekharbook} $(X_B,Y_B)$   and de Vries \citep{2000CQGdeVries} $(X_D,Y_D)$,  used  for computing the critical curve of Kerr-like black holes.  Although $(X_B,Y_B)$ and $(X_D,Y_D)$ are defined differently, they yield the same critical curve on the observer’s sky plane for a  stationary observer located at a large distance.    We find that $(X_B,Y_B)$  (refer  \cref{finite bardeen chandrasekhar X,finite bardeen chandrasekhar Y}) and $(X_D,Y_D)$  (refer  \cref{finite De Vries X,finite De Vries Y}) can yield different critical curves when $|p^{(t)}|\neq |p^{(r)}|$, which is significant if the observer is near the horizon or if the black hole is surrounded by a homogeneous plasma. We therefore compare $(X_B,Y_B)$ and $(X_D,Y_D)$ with the method of Grenzebach et al. \citep{Grenzebach:2014fha,Grenzebach2015book}, which computes the critical curve  using celestial angles $\gamma$ and $\delta$,  and  maps them to the observer’s sky plane using stereographic projection $(X_G,Y_G)$ (refer  \cref{Grenzebach X,Grenzebach Y}).  Throughout this work, we consider a stationary  observer in an axisymmetric, asymptotically flat Kerr-like spacetime.

  We express  the impact parameters  in terms of the spherical photon orbit radius $r_{sp}$,  using light ray geodesic equations for a black hole surrounded by a pressureless, non-magnetised plasma satisfying the condition \cref{separable plasma condition} \citep{Perlick:2017fio}.     We  set up an orthonormal tetrad for a stationary observer \citep{1993GReGrO} at a distance $D$ and inclination $\theta_i$ \citep{1993GReGrO}, and compute the local momenta.   For the  Kerr-like spacetime, we express  $(X_B,Y_B)$ and $(X_D,Y_D)$ in terms of tetrads $e^\mu_{(a)}$ while also using the geodesic equations in \cref{XB any tetrad,YB any tetrad,XD any tetrad,YD any tetrad}.

    We were also able to express $(X_B,Y_B)$ and $(X_D,Y_D)$ in terms of celestial angles $\gamma$ and $\delta$  in \cref{finite XB gamma delta,finite YB gamma delta,finite XD in gamma delta,finite YD in gamma delta}. This allowed for a  direct comparison  of $(X_B,Y_B)$, $(X_D,Y_D)$ with $(X_G,Y_G)$.   We report the following results obtained from the comparison:
\begin{itemize}
    \item  For a stationary observer at a finite distance, the equations of critical curves obtained using the definitions $(X_B,Y_B)$, $(X_D,Y_D)$ and $(X_G,Y_G)$ are different. 
    \item   By comparing $(X_B,Y_B)$  in \cref{finite XB gamma delta,finite YB gamma delta}, $(X_D,Y_D)$ in \cref{finite XD in gamma delta,finite YD in gamma delta}  and $(X_G,Y_G)$ in \cref{Grenzebach X,Grenzebach Y}, we were able to  analytically deduce that among the critical curves obtained using the three approaches, the  size of the critical curve computed using $(X_B,Y_B)$  will be the smallest, and $(X_D,Y_D)$ will be the largest.
    \item       When the observer distance is very large and plasma is absent $(\omega^2_P=0)$, then $(X_B,Y_B)$, $(X_D,Y_D)$ and $(X_G,Y_G)$   describe the same critical curve.  This is also true for a distant observer if  inhomogeneous plasma is considered, because at large distances for inhomogeneous plasma $\omega^2_P\rightarrow0$. Interestingly,  if the plasma is homogeneous, only $(X_D,Y_D)$ and $(X_G,Y_G)$  will produce the same critical curve at a large observer distance. 
    \item       More importantly, we show that the size of the critical curve computed using Bardeen's definition $(X_B,Y_B)$ decreases as the plasma density increases if the black hole is surrounded by homogeneous plasma. This effect persists even at a large distance.  This is in clear contradiction  with findings reported in  previous studies, which have shown that the critical curve size increases as the plasma density of homogeneous plasma increases.  This   contradictory feature of $(X_B,Y_B)$ has not been reported earlier, to the best of our knowledge.   Thus, in general, Bardeen's definition $(X_B,Y_B)$ \citep{Bardeen:1973,Chandrasekharbook}  is not suitable  for computing the critical curve of Kerr-like black holes. 
\end{itemize}
 The mismatch between $(X_D,Y_D)$ and $(X_G,Y_G)$ was resolved by   showing that  de Vries's definition \citep{2000CQGdeVries} $(X_D,Y_D)$  can be derived assuming a critical curve on the observer's sky plane. In other words, we were able to show that a point  on the critical curve can be expressed in terms of $(X_D,Y_D)$ as discussed in   \cref{Derivation of de Vries definition from the  critical curve}.  Thus, for an observer at a finite distance,  de Vries's definition \citep{2000CQGdeVries} $(X_D,Y_D)$  is suitable for computing/plotting the critical curve on the observer's sky plane.  We also identified the  main reason behind the mismatch between $(X_G,Y_G)$ and $(X_D,Y_D)$. The mismatch is primarily because of  the use of stereographic projection to trace the critical curve computed on the observer's sky plane  by Grenzebach et al. \citep{Grenzebach:2014fha,Grenzebach2015book}.  In other words,  for a finite distance observer, the point on the observer's sky plane  obtained from stereographic projection does not correctly represent the direction of the tangent vector of the light ray originating from the observer's position.   Note that all these results were obtained  assuming the observer and the orthonormal tetrads are stationary for an asymptotically flat Kerr-like spacetime (which has been widely adopted in the literature) and without using the separability condition of the plasma.

    We then consider the case of a Kerr black hole surrounded by a non-magnetised, pressureless plasma satisfying the separability condition \cref{separable plasma condition}. We obtain the equation of the critical curve  using $(X_B,Y_B)$, $(X_D,Y_D)$ and $(X_G,Y_G)$  with Bardeen tetrads and Carter tetrads in  \cref{Analysing the critical curve using Bardeen tetrad,Analysing the critical curve using Carter tetrad}, respectively. Using each of these tetrads, we demonstrate   the differences between $(X_B,Y_B)$, $(X_D,Y_D)$ and $(X_G,Y_G)$ for Schwarzschild and near extremal Kerr black holes at varying observer distances  and in the presence of both  homogeneous   and inhomogeneous plasma \citep{shapiro1974accretion,Perlick:2017fio}. 
    We report the following additional observations:
\begin{itemize}
    \item Effect of tetrad change  on critical curve: Switching from Bardeen to 
    Carter tetrads introduces a horizontal shift in the critical curve for a Kerr black hole (due to the combined effect of spin $a$ and observer inclination 
    $\theta_i$), while no such shift occurs for a Schwarzschild black hole, since both   tetrads reduce to the static tetrad in the limit $a \to 0$.
    \item  Additionally, the geometric radius $R_S$ of the Schwarzschild critical curve $(R_S=D\tan\gamma_S)$ as discussed by Synge \citep{Synge:1966okc} and Zel'dovich \& Novikov  \citep{Zeldovich1966}  is correctly     recovered only through  de Vries's definition. The Bardeen and Grenzebach et al.    definitions yield $D\sin\gamma_S$ and $2D\tan(\gamma_S/2)$, respectively, which   differ from $R_S$ significantly for a finite distance observer.
\end{itemize}

These results confirm that the inconsistencies between the three definitions are  generic and are not artefacts of a particular tetrad choice. The homogeneous plasma case most starkly exposes the contradiction between Bardeen's definition and the other two approaches, both at finite and astrophysical observer distances.
\bibliography{0reference}
\bibliographystyle{utphys1.bst} 
\end{document}